\documentclass[prd,superscriptaddress,twocolumn,nofootinbib,showpacs,preprintnumbers,floatfix]{revtex4}

\usepackage{mathrsfs}
\usepackage{amsfonts,amssymb,amsmath}
\usepackage{graphicx,epsfig}
\usepackage{multirow}
\usepackage{cancel}
\usepackage{verbatim}
\usepackage{color}
\definecolor{AliceBlue}{rgb}{0.94,0.97,1.00}
\definecolor{AntiqueWhite1}{rgb}{1.00,0.94,0.86}
\definecolor{AntiqueWhite2}{rgb}{0.93,0.87,0.80}
\definecolor{AntiqueWhite3}{rgb}{0.80,0.75,0.69}
\definecolor{AntiqueWhite4}{rgb}{0.55,0.51,0.47}
\definecolor{AntiqueWhite}{rgb}{0.98,0.92,0.84}
\definecolor{BlanchedAlmond}{rgb}{1.00,0.92,0.80}
\definecolor{BlueViolet}{rgb}{0.54,0.17,0.89}
\definecolor{CadetBlue1}{rgb}{0.60,0.96,1.00}
\definecolor{CadetBlue2}{rgb}{0.56,0.90,0.93}
\definecolor{CadetBlue3}{rgb}{0.48,0.77,0.80}
\definecolor{CadetBlue4}{rgb}{0.33,0.53,0.55}
\definecolor{CadetBlue}{rgb}{0.37,0.62,0.63}
\definecolor{CornflowerBlue}{rgb}{0.39,0.58,0.93}
\definecolor{DarkBlue}{rgb}{0.00,0.00,0.55}
\definecolor{DarkCyan}{rgb}{0.00,0.55,0.55}
\definecolor{DarkGoldenrod1}{rgb}{1.00,0.73,0.06}
\definecolor{DarkGoldenrod2}{rgb}{0.93,0.68,0.05}
\definecolor{DarkGoldenrod3}{rgb}{0.80,0.58,0.05}
\definecolor{DarkGoldenrod4}{rgb}{0.55,0.40,0.03}
\definecolor{DarkGoldenrod}{rgb}{0.72,0.53,0.04}
\definecolor{DarkGray}{rgb}{0.66,0.66,0.66}
\definecolor{DarkGreen}{rgb}{0.00,0.39,0.00}
\definecolor{DarkGrey}{rgb}{0.66,0.66,0.66}
\definecolor{DarkKhaki}{rgb}{0.74,0.72,0.42}
\definecolor{DarkMagenta}{rgb}{0.55,0.00,0.55}
\definecolor{DarkOliveGreen1}{rgb}{0.79,1.00,0.44}
\definecolor{DarkOliveGreen2}{rgb}{0.74,0.93,0.41}
\definecolor{DarkOliveGreen3}{rgb}{0.64,0.80,0.35}
\definecolor{DarkOliveGreen4}{rgb}{0.43,0.55,0.24}
\definecolor{DarkOliveGreen}{rgb}{0.33,0.42,0.18}
\definecolor{DarkOrange1}{rgb}{1.00,0.50,0.00}
\definecolor{DarkOrange2}{rgb}{0.93,0.46,0.00}
\definecolor{DarkOrange3}{rgb}{0.80,0.40,0.00}
\definecolor{DarkOrange4}{rgb}{0.55,0.27,0.00}
\definecolor{DarkOrange}{rgb}{1.00,0.55,0.00}
\definecolor{DarkOrchid1}{rgb}{0.75,0.24,1.00}
\definecolor{DarkOrchid2}{rgb}{0.70,0.23,0.93}
\definecolor{DarkOrchid3}{rgb}{0.60,0.20,0.80}
\definecolor{DarkOrchid4}{rgb}{0.41,0.13,0.55}
\definecolor{DarkOrchid}{rgb}{0.60,0.20,0.80}
\definecolor{DarkRed}{rgb}{0.55,0.00,0.00}
\definecolor{DarkSalmon}{rgb}{0.91,0.59,0.48}
\definecolor{DarkSeaGreen1}{rgb}{0.76,1.00,0.76}
\definecolor{DarkSeaGreen2}{rgb}{0.71,0.93,0.71}
\definecolor{DarkSeaGreen3}{rgb}{0.61,0.80,0.61}
\definecolor{DarkSeaGreen4}{rgb}{0.41,0.55,0.41}
\definecolor{DarkSeaGreen}{rgb}{0.56,0.74,0.56}
\definecolor{DarkSlateBlue}{rgb}{0.28,0.24,0.55}
\definecolor{DarkSlateGray1}{rgb}{0.59,1.00,1.00}
\definecolor{DarkSlateGray2}{rgb}{0.55,0.93,0.93}
\definecolor{DarkSlateGray3}{rgb}{0.47,0.80,0.80}
\definecolor{DarkSlateGray4}{rgb}{0.32,0.55,0.55}
\definecolor{DarkSlateGray}{rgb}{0.18,0.31,0.31}
\definecolor{DarkSlateGrey}{rgb}{0.18,0.31,0.31}
\definecolor{DarkTurquoise}{rgb}{0.00,0.81,0.82}
\definecolor{DarkViolet}{rgb}{0.58,0.00,0.83}
\definecolor{DeepPink1}{rgb}{1.00,0.08,0.58}
\definecolor{DeepPink2}{rgb}{0.93,0.07,0.54}
\definecolor{DeepPink3}{rgb}{0.80,0.06,0.46}
\definecolor{DeepPink4}{rgb}{0.55,0.04,0.31}
\definecolor{DeepPink}{rgb}{1.00,0.08,0.58}
\definecolor{DeepSkyBlue1}{rgb}{0.00,0.75,1.00}
\definecolor{DeepSkyBlue2}{rgb}{0.00,0.70,0.93}
\definecolor{DeepSkyBlue3}{rgb}{0.00,0.60,0.80}
\definecolor{DeepSkyBlue4}{rgb}{0.00,0.41,0.55}
\definecolor{DeepSkyBlue}{rgb}{0.00,0.75,1.00}
\definecolor{DimGray}{rgb}{0.41,0.41,0.41}
\definecolor{DimGrey}{rgb}{0.41,0.41,0.41}
\definecolor{DodgerBlue1}{rgb}{0.12,0.56,1.00}
\definecolor{DodgerBlue2}{rgb}{0.11,0.53,0.93}
\definecolor{DodgerBlue3}{rgb}{0.09,0.45,0.80}
\definecolor{DodgerBlue4}{rgb}{0.06,0.31,0.55}
\definecolor{DodgerBlue}{rgb}{0.12,0.56,1.00}
\definecolor{FloralWhite}{rgb}{1.00,0.98,0.94}
\definecolor{ForestGreen}{rgb}{0.13,0.55,0.13}
\definecolor{GhostWhite}{rgb}{0.97,0.97,1.00}
\definecolor{GreenYellow}{rgb}{0.68,1.00,0.18}
\definecolor{HotPink1}{rgb}{1.00,0.43,0.71}
\definecolor{HotPink2}{rgb}{0.93,0.42,0.65}
\definecolor{HotPink3}{rgb}{0.80,0.38,0.56}
\definecolor{HotPink4}{rgb}{0.55,0.23,0.38}
\definecolor{HotPink}{rgb}{1.00,0.41,0.71}
\definecolor{IndianRed1}{rgb}{1.00,0.42,0.42}
\definecolor{IndianRed2}{rgb}{0.93,0.39,0.39}
\definecolor{IndianRed3}{rgb}{0.80,0.33,0.33}
\definecolor{IndianRed4}{rgb}{0.55,0.23,0.23}
\definecolor{IndianRed}{rgb}{0.80,0.36,0.36}
\definecolor{LavenderBlush1}{rgb}{1.00,0.94,0.96}
\definecolor{LavenderBlush2}{rgb}{0.93,0.88,0.90}
\definecolor{LavenderBlush3}{rgb}{0.80,0.76,0.77}
\definecolor{LavenderBlush4}{rgb}{0.55,0.51,0.53}
\definecolor{LavenderBlush}{rgb}{1.00,0.94,0.96}
\definecolor{LawnGreen}{rgb}{0.49,0.99,0.00}
\definecolor{LemonChiffon1}{rgb}{1.00,0.98,0.80}
\definecolor{LemonChiffon2}{rgb}{0.93,0.91,0.75}
\definecolor{LemonChiffon3}{rgb}{0.80,0.79,0.65}
\definecolor{LemonChiffon4}{rgb}{0.55,0.54,0.44}
\definecolor{LemonChiffon}{rgb}{1.00,0.98,0.80}
\definecolor{LightBlue1}{rgb}{0.75,0.94,1.00}
\definecolor{LightBlue2}{rgb}{0.70,0.87,0.93}
\definecolor{LightBlue3}{rgb}{0.60,0.75,0.80}
\definecolor{LightBlue4}{rgb}{0.41,0.51,0.55}
\definecolor{LightBlue}{rgb}{0.68,0.85,0.90}
\definecolor{LightCoral}{rgb}{0.94,0.50,0.50}
\definecolor{LightCyan1}{rgb}{0.88,1.00,1.00}
\definecolor{LightCyan2}{rgb}{0.82,0.93,0.93}
\definecolor{LightCyan3}{rgb}{0.71,0.80,0.80}
\definecolor{LightCyan4}{rgb}{0.48,0.55,0.55}
\definecolor{LightCyan}{rgb}{0.88,1.00,1.00}
\definecolor{LightGoldenrod1}{rgb}{1.00,0.93,0.55}
\definecolor{LightGoldenrod2}{rgb}{0.93,0.86,0.51}
\definecolor{LightGoldenrod3}{rgb}{0.80,0.75,0.44}
\definecolor{LightGoldenrod4}{rgb}{0.55,0.51,0.30}
\definecolor{LightGoldenrodYellow}{rgb}{0.98,0.98,0.82}
\definecolor{LightGoldenrod}{rgb}{0.93,0.87,0.51}
\definecolor{LightGray}{rgb}{0.83,0.83,0.83}
\definecolor{LightGreen}{rgb}{0.56,0.93,0.56}
\definecolor{LightGrey}{rgb}{0.83,0.83,0.83}
\definecolor{LightPink1}{rgb}{1.00,0.68,0.73}
\definecolor{LightPink2}{rgb}{0.93,0.64,0.68}
\definecolor{LightPink3}{rgb}{0.80,0.55,0.58}
\definecolor{LightPink4}{rgb}{0.55,0.37,0.40}
\definecolor{LightPink}{rgb}{1.00,0.71,0.76}
\definecolor{LightSalmon1}{rgb}{1.00,0.63,0.48}
\definecolor{LightSalmon2}{rgb}{0.93,0.58,0.45}
\definecolor{LightSalmon3}{rgb}{0.80,0.51,0.38}
\definecolor{LightSalmon4}{rgb}{0.55,0.34,0.26}
\definecolor{LightSalmon}{rgb}{1.00,0.63,0.48}
\definecolor{LightSeaGreen}{rgb}{0.13,0.70,0.67}
\definecolor{LightSkyBlue1}{rgb}{0.69,0.89,1.00}
\definecolor{LightSkyBlue2}{rgb}{0.64,0.83,0.93}
\definecolor{LightSkyBlue3}{rgb}{0.55,0.71,0.80}
\definecolor{LightSkyBlue4}{rgb}{0.38,0.48,0.55}
\definecolor{LightSkyBlue}{rgb}{0.53,0.81,0.98}
\definecolor{LightSlateBlue}{rgb}{0.52,0.44,1.00}
\definecolor{LightSlateGray}{rgb}{0.47,0.53,0.60}
\definecolor{LightSlateGrey}{rgb}{0.47,0.53,0.60}
\definecolor{LightSteelBlue1}{rgb}{0.79,0.88,1.00}
\definecolor{LightSteelBlue2}{rgb}{0.74,0.82,0.93}
\definecolor{LightSteelBlue3}{rgb}{0.64,0.71,0.80}
\definecolor{LightSteelBlue4}{rgb}{0.43,0.48,0.55}
\definecolor{LightSteelBlue}{rgb}{0.69,0.77,0.87}
\definecolor{LightYellow1}{rgb}{1.00,1.00,0.88}
\definecolor{LightYellow2}{rgb}{0.93,0.93,0.82}
\definecolor{LightYellow3}{rgb}{0.80,0.80,0.71}
\definecolor{LightYellow4}{rgb}{0.55,0.55,0.48}
\definecolor{LightYellow}{rgb}{1.00,1.00,0.88}
\definecolor{LimeGreen}{rgb}{0.20,0.80,0.20}
\definecolor{MediumAquamarine}{rgb}{0.40,0.80,0.67}
\definecolor{MediumBlue}{rgb}{0.00,0.00,0.80}
\definecolor{MediumOrchid1}{rgb}{0.88,0.40,1.00}
\definecolor{MediumOrchid2}{rgb}{0.82,0.37,0.93}
\definecolor{MediumOrchid3}{rgb}{0.71,0.32,0.80}
\definecolor{MediumOrchid4}{rgb}{0.48,0.22,0.55}
\definecolor{MediumOrchid}{rgb}{0.73,0.33,0.83}
\definecolor{MediumPurple1}{rgb}{0.67,0.51,1.00}
\definecolor{MediumPurple2}{rgb}{0.62,0.47,0.93}
\definecolor{MediumPurple3}{rgb}{0.54,0.41,0.80}
\definecolor{MediumPurple4}{rgb}{0.36,0.28,0.55}
\definecolor{MediumPurple}{rgb}{0.58,0.44,0.86}
\definecolor{MediumSeaGreen}{rgb}{0.24,0.70,0.44}
\definecolor{MediumSlateBlue}{rgb}{0.48,0.41,0.93}
\definecolor{MediumSpringGreen}{rgb}{0.00,0.98,0.60}
\definecolor{MediumTurquoise}{rgb}{0.28,0.82,0.80}
\definecolor{MediumVioletRed}{rgb}{0.78,0.08,0.52}
\definecolor{MidnightBlue}{rgb}{0.10,0.10,0.44}
\definecolor{MintCream}{rgb}{0.96,1.00,0.98}
\definecolor{MistyRose1}{rgb}{1.00,0.89,0.88}
\definecolor{MistyRose2}{rgb}{0.93,0.84,0.82}
\definecolor{MistyRose3}{rgb}{0.80,0.72,0.71}
\definecolor{MistyRose4}{rgb}{0.55,0.49,0.48}
\definecolor{MistyRose}{rgb}{1.00,0.89,0.88}
\definecolor{NavajoWhite1}{rgb}{1.00,0.87,0.68}
\definecolor{NavajoWhite2}{rgb}{0.93,0.81,0.63}
\definecolor{NavajoWhite3}{rgb}{0.80,0.70,0.55}
\definecolor{NavajoWhite4}{rgb}{0.55,0.47,0.37}
\definecolor{NavajoWhite}{rgb}{1.00,0.87,0.68}
\definecolor{NavyBlue}{rgb}{0.00,0.00,0.50}
\definecolor{OldLace}{rgb}{0.99,0.96,0.90}
\definecolor{OliveDrab1}{rgb}{0.75,1.00,0.24}
\definecolor{OliveDrab2}{rgb}{0.70,0.93,0.23}
\definecolor{OliveDrab3}{rgb}{0.60,0.80,0.20}
\definecolor{OliveDrab4}{rgb}{0.41,0.55,0.13}
\definecolor{OliveDrab}{rgb}{0.42,0.56,0.14}
\definecolor{OrangeRed1}{rgb}{1.00,0.27,0.00}
\definecolor{OrangeRed2}{rgb}{0.93,0.25,0.00}
\definecolor{OrangeRed3}{rgb}{0.80,0.22,0.00}
\definecolor{OrangeRed4}{rgb}{0.55,0.15,0.00}
\definecolor{OrangeRed}{rgb}{1.00,0.27,0.00}
\definecolor{PaleGoldenrod}{rgb}{0.93,0.91,0.67}
\definecolor{PaleGreen1}{rgb}{0.60,1.00,0.60}
\definecolor{PaleGreen2}{rgb}{0.56,0.93,0.56}
\definecolor{PaleGreen3}{rgb}{0.49,0.80,0.49}
\definecolor{PaleGreen4}{rgb}{0.33,0.55,0.33}
\definecolor{PaleGreen}{rgb}{0.60,0.98,0.60}
\definecolor{PaleTurquoise1}{rgb}{0.73,1.00,1.00}
\definecolor{PaleTurquoise2}{rgb}{0.68,0.93,0.93}
\definecolor{PaleTurquoise3}{rgb}{0.59,0.80,0.80}
\definecolor{PaleTurquoise4}{rgb}{0.40,0.55,0.55}
\definecolor{PaleTurquoise}{rgb}{0.69,0.93,0.93}
\definecolor{PaleVioletRed1}{rgb}{1.00,0.51,0.67}
\definecolor{PaleVioletRed2}{rgb}{0.93,0.47,0.62}
\definecolor{PaleVioletRed3}{rgb}{0.80,0.41,0.54}
\definecolor{PaleVioletRed4}{rgb}{0.55,0.28,0.36}
\definecolor{PaleVioletRed}{rgb}{0.86,0.44,0.58}
\definecolor{PapayaWhip}{rgb}{1.00,0.94,0.84}
\definecolor{PeachPuff1}{rgb}{1.00,0.85,0.73}
\definecolor{PeachPuff2}{rgb}{0.93,0.80,0.68}
\definecolor{PeachPuff3}{rgb}{0.80,0.69,0.58}
\definecolor{PeachPuff4}{rgb}{0.55,0.47,0.40}
\definecolor{PeachPuff}{rgb}{1.00,0.85,0.73}
\definecolor{PowderBlue}{rgb}{0.69,0.88,0.90}
\definecolor{RosyBrown1}{rgb}{1.00,0.76,0.76}
\definecolor{RosyBrown2}{rgb}{0.93,0.71,0.71}
\definecolor{RosyBrown3}{rgb}{0.80,0.61,0.61}
\definecolor{RosyBrown4}{rgb}{0.55,0.41,0.41}
\definecolor{RosyBrown}{rgb}{0.74,0.56,0.56}
\definecolor{RoyalBlue1}{rgb}{0.28,0.46,1.00}
\definecolor{RoyalBlue2}{rgb}{0.26,0.43,0.93}
\definecolor{RoyalBlue3}{rgb}{0.23,0.37,0.80}
\definecolor{RoyalBlue4}{rgb}{0.15,0.25,0.55}
\definecolor{RoyalBlue}{rgb}{0.25,0.41,0.88}
\definecolor{SaddleBrown}{rgb}{0.55,0.27,0.07}
\definecolor{SandyBrown}{rgb}{0.96,0.64,0.38}
\definecolor{SeaGreen1}{rgb}{0.33,1.00,0.62}
\definecolor{SeaGreen2}{rgb}{0.31,0.93,0.58}
\definecolor{SeaGreen3}{rgb}{0.26,0.80,0.50}
\definecolor{SeaGreen4}{rgb}{0.18,0.55,0.34}
\definecolor{SeaGreen}{rgb}{0.18,0.55,0.34}
\definecolor{SkyBlue1}{rgb}{0.53,0.81,1.00}
\definecolor{SkyBlue2}{rgb}{0.49,0.75,0.93}
\definecolor{SkyBlue3}{rgb}{0.42,0.65,0.80}
\definecolor{SkyBlue4}{rgb}{0.29,0.44,0.55}
\definecolor{SkyBlue}{rgb}{0.53,0.81,0.92}
\definecolor{SlateBlue1}{rgb}{0.51,0.44,1.00}
\definecolor{SlateBlue2}{rgb}{0.48,0.40,0.93}
\definecolor{SlateBlue3}{rgb}{0.41,0.35,0.80}
\definecolor{SlateBlue4}{rgb}{0.28,0.24,0.55}
\definecolor{SlateBlue}{rgb}{0.42,0.35,0.80}
\definecolor{SlateGray1}{rgb}{0.78,0.89,1.00}
\definecolor{SlateGray2}{rgb}{0.73,0.83,0.93}
\definecolor{SlateGray3}{rgb}{0.62,0.71,0.80}
\definecolor{SlateGray4}{rgb}{0.42,0.48,0.55}
\definecolor{SlateGray}{rgb}{0.44,0.50,0.56}
\definecolor{SlateGrey}{rgb}{0.44,0.50,0.56}
\definecolor{SpringGreen1}{rgb}{0.00,1.00,0.50}
\definecolor{SpringGreen2}{rgb}{0.00,0.93,0.46}
\definecolor{SpringGreen3}{rgb}{0.00,0.80,0.40}
\definecolor{SpringGreen4}{rgb}{0.00,0.55,0.27}
\definecolor{SpringGreen}{rgb}{0.00,1.00,0.50}
\definecolor{SteelBlue1}{rgb}{0.39,0.72,1.00}
\definecolor{SteelBlue2}{rgb}{0.36,0.67,0.93}
\definecolor{SteelBlue3}{rgb}{0.31,0.58,0.80}
\definecolor{SteelBlue4}{rgb}{0.21,0.39,0.55}
\definecolor{SteelBlue}{rgb}{0.27,0.51,0.71}
\definecolor{VioletRed1}{rgb}{1.00,0.24,0.59}
\definecolor{VioletRed2}{rgb}{0.93,0.23,0.55}
\definecolor{VioletRed3}{rgb}{0.80,0.20,0.47}
\definecolor{VioletRed4}{rgb}{0.55,0.13,0.32}
\definecolor{VioletRed}{rgb}{0.82,0.13,0.56}
\definecolor{WhiteSmoke}{rgb}{0.96,0.96,0.96}
\definecolor{YellowGreen}{rgb}{0.60,0.80,0.20}
\definecolor{aliceblue}{rgb}{0.94,0.97,1.00}
\definecolor{antiquewhite}{rgb}{0.98,0.92,0.84}
\definecolor{aquamarine1}{rgb}{0.50,1.00,0.83}
\definecolor{aquamarine2}{rgb}{0.46,0.93,0.78}
\definecolor{aquamarine3}{rgb}{0.40,0.80,0.67}
\definecolor{aquamarine4}{rgb}{0.27,0.55,0.45}
\definecolor{aquamarine}{rgb}{0.50,1.00,0.83}
\definecolor{azure1}{rgb}{0.94,1.00,1.00}
\definecolor{azure2}{rgb}{0.88,0.93,0.93}
\definecolor{azure3}{rgb}{0.76,0.80,0.80}
\definecolor{azure4}{rgb}{0.51,0.55,0.55}
\definecolor{azure}{rgb}{0.94,1.00,1.00}
\definecolor{beige}{rgb}{0.96,0.96,0.86}
\definecolor{bisque1}{rgb}{1.00,0.89,0.77}
\definecolor{bisque2}{rgb}{0.93,0.84,0.72}
\definecolor{bisque3}{rgb}{0.80,0.72,0.62}
\definecolor{bisque4}{rgb}{0.55,0.49,0.42}
\definecolor{bisque}{rgb}{1.00,0.89,0.77}
\definecolor{black}{rgb}{0.00,0.00,0.00}
\definecolor{blanchedalmond}{rgb}{1.00,0.92,0.80}
\definecolor{blue1}{rgb}{0.00,0.00,1.00}
\definecolor{blue2}{rgb}{0.00,0.00,0.93}
\definecolor{blue3}{rgb}{0.00,0.00,0.80}
\definecolor{blue4}{rgb}{0.00,0.00,0.55}
\definecolor{blueviolet}{rgb}{0.54,0.17,0.89}
\definecolor{blue}{rgb}{0.00,0.00,1.00}
\definecolor{brown1}{rgb}{1.00,0.25,0.25}
\definecolor{brown2}{rgb}{0.93,0.23,0.23}
\definecolor{brown3}{rgb}{0.80,0.20,0.20}
\definecolor{brown4}{rgb}{0.55,0.14,0.14}
\definecolor{brown}{rgb}{0.65,0.16,0.16}
\definecolor{burlywood1}{rgb}{1.00,0.83,0.61}
\definecolor{burlywood2}{rgb}{0.93,0.77,0.57}
\definecolor{burlywood3}{rgb}{0.80,0.67,0.49}
\definecolor{burlywood4}{rgb}{0.55,0.45,0.33}
\definecolor{burlywood}{rgb}{0.87,0.72,0.53}
\definecolor{cadetblue}{rgb}{0.37,0.62,0.63}
\definecolor{chartreuse1}{rgb}{0.50,1.00,0.00}
\definecolor{chartreuse2}{rgb}{0.46,0.93,0.00}
\definecolor{chartreuse3}{rgb}{0.40,0.80,0.00}
\definecolor{chartreuse4}{rgb}{0.27,0.55,0.00}
\definecolor{chartreuse}{rgb}{0.50,1.00,0.00}
\definecolor{chocolate1}{rgb}{1.00,0.50,0.14}
\definecolor{chocolate2}{rgb}{0.93,0.46,0.13}
\definecolor{chocolate3}{rgb}{0.80,0.40,0.11}
\definecolor{chocolate4}{rgb}{0.55,0.27,0.07}
\definecolor{chocolate}{rgb}{0.82,0.41,0.12}
\definecolor{coral1}{rgb}{1.00,0.45,0.34}
\definecolor{coral2}{rgb}{0.93,0.42,0.31}
\definecolor{coral3}{rgb}{0.80,0.36,0.27}
\definecolor{coral4}{rgb}{0.55,0.24,0.18}
\definecolor{coral}{rgb}{1.00,0.50,0.31}
\definecolor{cornflowerblue}{rgb}{0.39,0.58,0.93}
\definecolor{cornsilk1}{rgb}{1.00,0.97,0.86}
\definecolor{cornsilk2}{rgb}{0.93,0.91,0.80}
\definecolor{cornsilk3}{rgb}{0.80,0.78,0.69}
\definecolor{cornsilk4}{rgb}{0.55,0.53,0.47}
\definecolor{cornsilk}{rgb}{1.00,0.97,0.86}
\definecolor{cyan1}{rgb}{0.00,1.00,1.00}
\definecolor{cyan2}{rgb}{0.00,0.93,0.93}
\definecolor{cyan3}{rgb}{0.00,0.80,0.80}
\definecolor{cyan4}{rgb}{0.00,0.55,0.55}
\definecolor{cyan}{rgb}{0.00,1.00,1.00}
\definecolor{darkblue}{rgb}{0.00,0.00,0.55}
\definecolor{darkcyan}{rgb}{0.00,0.55,0.55}
\definecolor{darkgoldenrod}{rgb}{0.72,0.53,0.04}
\definecolor{darkgray}{rgb}{0.66,0.66,0.66}
\definecolor{darkgreen}{rgb}{0.00,0.39,0.00}
\definecolor{darkgrey}{rgb}{0.66,0.66,0.66}
\definecolor{darkkhaki}{rgb}{0.74,0.72,0.42}
\definecolor{darkmagenta}{rgb}{0.55,0.00,0.55}
\definecolor{darkolive}{rgb}{0.33,0.42,0.18}
\definecolor{darkorange}{rgb}{1.00,0.55,0.00}
\definecolor{darkorchid}{rgb}{0.60,0.20,0.80}
\definecolor{darkred}{rgb}{0.55,0.00,0.00}
\definecolor{darksalmon}{rgb}{0.91,0.59,0.48}
\definecolor{darksea}{rgb}{0.56,0.74,0.56}
\definecolor{darkslate}{rgb}{0.18,0.31,0.31}
\definecolor{darkslate}{rgb}{0.18,0.31,0.31}
\definecolor{darkslate}{rgb}{0.28,0.24,0.55}
\definecolor{darkturquoise}{rgb}{0.00,0.81,0.82}
\definecolor{darkviolet}{rgb}{0.58,0.00,0.83}
\definecolor{deeppink}{rgb}{1.00,0.08,0.58}
\definecolor{deepsky}{rgb}{0.00,0.75,1.00}
\definecolor{dimgray}{rgb}{0.41,0.41,0.41}
\definecolor{dimgrey}{rgb}{0.41,0.41,0.41}
\definecolor{dodgerblue}{rgb}{0.12,0.56,1.00}
\definecolor{firebrick1}{rgb}{1.00,0.19,0.19}
\definecolor{firebrick2}{rgb}{0.93,0.17,0.17}
\definecolor{firebrick3}{rgb}{0.80,0.15,0.15}
\definecolor{firebrick4}{rgb}{0.55,0.10,0.10}
\definecolor{firebrick}{rgb}{0.70,0.13,0.13}
\definecolor{floralwhite}{rgb}{1.00,0.98,0.94}
\definecolor{forestgreen}{rgb}{0.13,0.55,0.13}
\definecolor{gainsboro}{rgb}{0.86,0.86,0.86}
\definecolor{ghostwhite}{rgb}{0.97,0.97,1.00}
\definecolor{gold1}{rgb}{1.00,0.84,0.00}
\definecolor{gold2}{rgb}{0.93,0.79,0.00}
\definecolor{gold3}{rgb}{0.80,0.68,0.00}
\definecolor{gold4}{rgb}{0.55,0.46,0.00}
\definecolor{goldenrod1}{rgb}{1.00,0.76,0.15}
\definecolor{goldenrod2}{rgb}{0.93,0.71,0.13}
\definecolor{goldenrod3}{rgb}{0.80,0.61,0.11}
\definecolor{goldenrod4}{rgb}{0.55,0.41,0.08}
\definecolor{goldenrod}{rgb}{0.85,0.65,0.13}
\definecolor{gold}{rgb}{1.00,0.84,0.00}
\definecolor{gray0}{rgb}{0.00,0.00,0.00}
\definecolor{gray100}{rgb}{1.00,1.00,1.00}
\definecolor{gray10}{rgb}{0.10,0.10,0.10}
\definecolor{gray11}{rgb}{0.11,0.11,0.11}
\definecolor{gray12}{rgb}{0.12,0.12,0.12}
\definecolor{gray13}{rgb}{0.13,0.13,0.13}
\definecolor{gray14}{rgb}{0.14,0.14,0.14}
\definecolor{gray15}{rgb}{0.15,0.15,0.15}
\definecolor{gray16}{rgb}{0.16,0.16,0.16}
\definecolor{gray17}{rgb}{0.17,0.17,0.17}
\definecolor{gray18}{rgb}{0.18,0.18,0.18}
\definecolor{gray19}{rgb}{0.19,0.19,0.19}
\definecolor{gray1}{rgb}{0.01,0.01,0.01}
\definecolor{gray20}{rgb}{0.20,0.20,0.20}
\definecolor{gray21}{rgb}{0.21,0.21,0.21}
\definecolor{gray22}{rgb}{0.22,0.22,0.22}
\definecolor{gray23}{rgb}{0.23,0.23,0.23}
\definecolor{gray24}{rgb}{0.24,0.24,0.24}
\definecolor{gray25}{rgb}{0.25,0.25,0.25}
\definecolor{gray26}{rgb}{0.26,0.26,0.26}
\definecolor{gray27}{rgb}{0.27,0.27,0.27}
\definecolor{gray28}{rgb}{0.28,0.28,0.28}
\definecolor{gray29}{rgb}{0.29,0.29,0.29}
\definecolor{gray2}{rgb}{0.02,0.02,0.02}
\definecolor{gray30}{rgb}{0.30,0.30,0.30}
\definecolor{gray31}{rgb}{0.31,0.31,0.31}
\definecolor{gray32}{rgb}{0.32,0.32,0.32}
\definecolor{gray33}{rgb}{0.33,0.33,0.33}
\definecolor{gray34}{rgb}{0.34,0.34,0.34}
\definecolor{gray35}{rgb}{0.35,0.35,0.35}
\definecolor{gray36}{rgb}{0.36,0.36,0.36}
\definecolor{gray37}{rgb}{0.37,0.37,0.37}
\definecolor{gray38}{rgb}{0.38,0.38,0.38}
\definecolor{gray39}{rgb}{0.39,0.39,0.39}
\definecolor{gray3}{rgb}{0.03,0.03,0.03}
\definecolor{gray40}{rgb}{0.40,0.40,0.40}
\definecolor{gray41}{rgb}{0.41,0.41,0.41}
\definecolor{gray42}{rgb}{0.42,0.42,0.42}
\definecolor{gray43}{rgb}{0.43,0.43,0.43}
\definecolor{gray44}{rgb}{0.44,0.44,0.44}
\definecolor{gray45}{rgb}{0.45,0.45,0.45}
\definecolor{gray46}{rgb}{0.46,0.46,0.46}
\definecolor{gray47}{rgb}{0.47,0.47,0.47}
\definecolor{gray48}{rgb}{0.48,0.48,0.48}
\definecolor{gray49}{rgb}{0.49,0.49,0.49}
\definecolor{gray4}{rgb}{0.04,0.04,0.04}
\definecolor{gray50}{rgb}{0.50,0.50,0.50}
\definecolor{gray51}{rgb}{0.51,0.51,0.51}
\definecolor{gray52}{rgb}{0.52,0.52,0.52}
\definecolor{gray53}{rgb}{0.53,0.53,0.53}
\definecolor{gray54}{rgb}{0.54,0.54,0.54}
\definecolor{gray55}{rgb}{0.55,0.55,0.55}
\definecolor{gray56}{rgb}{0.56,0.56,0.56}
\definecolor{gray57}{rgb}{0.57,0.57,0.57}
\definecolor{gray58}{rgb}{0.58,0.58,0.58}
\definecolor{gray59}{rgb}{0.59,0.59,0.59}
\definecolor{gray5}{rgb}{0.05,0.05,0.05}
\definecolor{gray60}{rgb}{0.60,0.60,0.60}
\definecolor{gray61}{rgb}{0.61,0.61,0.61}
\definecolor{gray62}{rgb}{0.62,0.62,0.62}
\definecolor{gray63}{rgb}{0.63,0.63,0.63}
\definecolor{gray64}{rgb}{0.64,0.64,0.64}
\definecolor{gray65}{rgb}{0.65,0.65,0.65}
\definecolor{gray66}{rgb}{0.66,0.66,0.66}
\definecolor{gray67}{rgb}{0.67,0.67,0.67}
\definecolor{gray68}{rgb}{0.68,0.68,0.68}
\definecolor{gray69}{rgb}{0.69,0.69,0.69}
\definecolor{gray6}{rgb}{0.06,0.06,0.06}
\definecolor{gray70}{rgb}{0.70,0.70,0.70}
\definecolor{gray71}{rgb}{0.71,0.71,0.71}
\definecolor{gray72}{rgb}{0.72,0.72,0.72}
\definecolor{gray73}{rgb}{0.73,0.73,0.73}
\definecolor{gray74}{rgb}{0.74,0.74,0.74}
\definecolor{gray75}{rgb}{0.75,0.75,0.75}
\definecolor{gray76}{rgb}{0.76,0.76,0.76}
\definecolor{gray77}{rgb}{0.77,0.77,0.77}
\definecolor{gray78}{rgb}{0.78,0.78,0.78}
\definecolor{gray79}{rgb}{0.79,0.79,0.79}
\definecolor{gray7}{rgb}{0.07,0.07,0.07}
\definecolor{gray80}{rgb}{0.80,0.80,0.80}
\definecolor{gray81}{rgb}{0.81,0.81,0.81}
\definecolor{gray82}{rgb}{0.82,0.82,0.82}
\definecolor{gray83}{rgb}{0.83,0.83,0.83}
\definecolor{gray84}{rgb}{0.84,0.84,0.84}
\definecolor{gray85}{rgb}{0.85,0.85,0.85}
\definecolor{gray86}{rgb}{0.86,0.86,0.86}
\definecolor{gray87}{rgb}{0.87,0.87,0.87}
\definecolor{gray88}{rgb}{0.88,0.88,0.88}
\definecolor{gray89}{rgb}{0.89,0.89,0.89}
\definecolor{gray8}{rgb}{0.08,0.08,0.08}
\definecolor{gray90}{rgb}{0.90,0.90,0.90}
\definecolor{gray91}{rgb}{0.91,0.91,0.91}
\definecolor{gray92}{rgb}{0.92,0.92,0.92}
\definecolor{gray93}{rgb}{0.93,0.93,0.93}
\definecolor{gray94}{rgb}{0.94,0.94,0.94}
\definecolor{gray95}{rgb}{0.95,0.95,0.95}
\definecolor{gray96}{rgb}{0.96,0.96,0.96}
\definecolor{gray97}{rgb}{0.97,0.97,0.97}
\definecolor{gray98}{rgb}{0.98,0.98,0.98}
\definecolor{gray99}{rgb}{0.99,0.99,0.99}
\definecolor{gray9}{rgb}{0.09,0.09,0.09}
\definecolor{gray}{rgb}{0.75,0.75,0.75}
\definecolor{green1}{rgb}{0.00,1.00,0.00}
\definecolor{green2}{rgb}{0.00,0.93,0.00}
\definecolor{green3}{rgb}{0.00,0.80,0.00}
\definecolor{green4}{rgb}{0.00,0.55,0.00}
\definecolor{greenyellow}{rgb}{0.68,1.00,0.18}
\definecolor{green}{rgb}{0.00,1.00,0.00}
\definecolor{grey0}{rgb}{0.00,0.00,0.00}
\definecolor{grey100}{rgb}{1.00,1.00,1.00}
\definecolor{grey10}{rgb}{0.10,0.10,0.10}
\definecolor{grey11}{rgb}{0.11,0.11,0.11}
\definecolor{grey12}{rgb}{0.12,0.12,0.12}
\definecolor{grey13}{rgb}{0.13,0.13,0.13}
\definecolor{grey14}{rgb}{0.14,0.14,0.14}
\definecolor{grey15}{rgb}{0.15,0.15,0.15}
\definecolor{grey16}{rgb}{0.16,0.16,0.16}
\definecolor{grey17}{rgb}{0.17,0.17,0.17}
\definecolor{grey18}{rgb}{0.18,0.18,0.18}
\definecolor{grey19}{rgb}{0.19,0.19,0.19}
\definecolor{grey1}{rgb}{0.01,0.01,0.01}
\definecolor{grey20}{rgb}{0.20,0.20,0.20}
\definecolor{grey21}{rgb}{0.21,0.21,0.21}
\definecolor{grey22}{rgb}{0.22,0.22,0.22}
\definecolor{grey23}{rgb}{0.23,0.23,0.23}
\definecolor{grey24}{rgb}{0.24,0.24,0.24}
\definecolor{grey25}{rgb}{0.25,0.25,0.25}
\definecolor{grey26}{rgb}{0.26,0.26,0.26}
\definecolor{grey27}{rgb}{0.27,0.27,0.27}
\definecolor{grey28}{rgb}{0.28,0.28,0.28}
\definecolor{grey29}{rgb}{0.29,0.29,0.29}
\definecolor{grey2}{rgb}{0.02,0.02,0.02}
\definecolor{grey30}{rgb}{0.30,0.30,0.30}
\definecolor{grey31}{rgb}{0.31,0.31,0.31}
\definecolor{grey32}{rgb}{0.32,0.32,0.32}
\definecolor{grey33}{rgb}{0.33,0.33,0.33}
\definecolor{grey34}{rgb}{0.34,0.34,0.34}
\definecolor{grey35}{rgb}{0.35,0.35,0.35}
\definecolor{grey36}{rgb}{0.36,0.36,0.36}
\definecolor{grey37}{rgb}{0.37,0.37,0.37}
\definecolor{grey38}{rgb}{0.38,0.38,0.38}
\definecolor{grey39}{rgb}{0.39,0.39,0.39}
\definecolor{grey3}{rgb}{0.03,0.03,0.03}
\definecolor{grey40}{rgb}{0.40,0.40,0.40}
\definecolor{grey41}{rgb}{0.41,0.41,0.41}
\definecolor{grey42}{rgb}{0.42,0.42,0.42}
\definecolor{grey43}{rgb}{0.43,0.43,0.43}
\definecolor{grey44}{rgb}{0.44,0.44,0.44}
\definecolor{grey45}{rgb}{0.45,0.45,0.45}
\definecolor{grey46}{rgb}{0.46,0.46,0.46}
\definecolor{grey47}{rgb}{0.47,0.47,0.47}
\definecolor{grey48}{rgb}{0.48,0.48,0.48}
\definecolor{grey49}{rgb}{0.49,0.49,0.49}
\definecolor{grey4}{rgb}{0.04,0.04,0.04}
\definecolor{grey50}{rgb}{0.50,0.50,0.50}
\definecolor{grey51}{rgb}{0.51,0.51,0.51}
\definecolor{grey52}{rgb}{0.52,0.52,0.52}
\definecolor{grey53}{rgb}{0.53,0.53,0.53}
\definecolor{grey54}{rgb}{0.54,0.54,0.54}
\definecolor{grey55}{rgb}{0.55,0.55,0.55}
\definecolor{grey56}{rgb}{0.56,0.56,0.56}
\definecolor{grey57}{rgb}{0.57,0.57,0.57}
\definecolor{grey58}{rgb}{0.58,0.58,0.58}
\definecolor{grey59}{rgb}{0.59,0.59,0.59}
\definecolor{grey5}{rgb}{0.05,0.05,0.05}
\definecolor{grey60}{rgb}{0.60,0.60,0.60}
\definecolor{grey61}{rgb}{0.61,0.61,0.61}
\definecolor{grey62}{rgb}{0.62,0.62,0.62}
\definecolor{grey63}{rgb}{0.63,0.63,0.63}
\definecolor{grey64}{rgb}{0.64,0.64,0.64}
\definecolor{grey65}{rgb}{0.65,0.65,0.65}
\definecolor{grey66}{rgb}{0.66,0.66,0.66}
\definecolor{grey67}{rgb}{0.67,0.67,0.67}
\definecolor{grey68}{rgb}{0.68,0.68,0.68}
\definecolor{grey69}{rgb}{0.69,0.69,0.69}
\definecolor{grey6}{rgb}{0.06,0.06,0.06}
\definecolor{grey70}{rgb}{0.70,0.70,0.70}
\definecolor{grey71}{rgb}{0.71,0.71,0.71}
\definecolor{grey72}{rgb}{0.72,0.72,0.72}
\definecolor{grey73}{rgb}{0.73,0.73,0.73}
\definecolor{grey74}{rgb}{0.74,0.74,0.74}
\definecolor{grey75}{rgb}{0.75,0.75,0.75}
\definecolor{grey76}{rgb}{0.76,0.76,0.76}
\definecolor{grey77}{rgb}{0.77,0.77,0.77}
\definecolor{grey78}{rgb}{0.78,0.78,0.78}
\definecolor{grey79}{rgb}{0.79,0.79,0.79}
\definecolor{grey7}{rgb}{0.07,0.07,0.07}
\definecolor{grey80}{rgb}{0.80,0.80,0.80}
\definecolor{grey81}{rgb}{0.81,0.81,0.81}
\definecolor{grey82}{rgb}{0.82,0.82,0.82}
\definecolor{grey83}{rgb}{0.83,0.83,0.83}
\definecolor{grey84}{rgb}{0.84,0.84,0.84}
\definecolor{grey85}{rgb}{0.85,0.85,0.85}
\definecolor{grey86}{rgb}{0.86,0.86,0.86}
\definecolor{grey87}{rgb}{0.87,0.87,0.87}
\definecolor{grey88}{rgb}{0.88,0.88,0.88}
\definecolor{grey89}{rgb}{0.89,0.89,0.89}
\definecolor{grey8}{rgb}{0.08,0.08,0.08}
\definecolor{grey90}{rgb}{0.90,0.90,0.90}
\definecolor{grey91}{rgb}{0.91,0.91,0.91}
\definecolor{grey92}{rgb}{0.92,0.92,0.92}
\definecolor{grey93}{rgb}{0.93,0.93,0.93}
\definecolor{grey94}{rgb}{0.94,0.94,0.94}
\definecolor{grey95}{rgb}{0.95,0.95,0.95}
\definecolor{grey96}{rgb}{0.96,0.96,0.96}
\definecolor{grey97}{rgb}{0.97,0.97,0.97}
\definecolor{grey98}{rgb}{0.98,0.98,0.98}
\definecolor{grey99}{rgb}{0.99,0.99,0.99}
\definecolor{grey9}{rgb}{0.09,0.09,0.09}
\definecolor{grey}{rgb}{0.75,0.75,0.75}
\definecolor{honeydew1}{rgb}{0.94,1.00,0.94}
\definecolor{honeydew2}{rgb}{0.88,0.93,0.88}
\definecolor{honeydew3}{rgb}{0.76,0.80,0.76}
\definecolor{honeydew4}{rgb}{0.51,0.55,0.51}
\definecolor{honeydew}{rgb}{0.94,1.00,0.94}
\definecolor{hotpink}{rgb}{1.00,0.41,0.71}
\definecolor{indianred}{rgb}{0.80,0.36,0.36}
\definecolor{ivory1}{rgb}{1.00,1.00,0.94}
\definecolor{ivory2}{rgb}{0.93,0.93,0.88}
\definecolor{ivory3}{rgb}{0.80,0.80,0.76}
\definecolor{ivory4}{rgb}{0.55,0.55,0.51}
\definecolor{ivory}{rgb}{1.00,1.00,0.94}
\definecolor{khaki1}{rgb}{1.00,0.96,0.56}
\definecolor{khaki2}{rgb}{0.93,0.90,0.52}
\definecolor{khaki3}{rgb}{0.80,0.78,0.45}
\definecolor{khaki4}{rgb}{0.55,0.53,0.31}
\definecolor{khaki}{rgb}{0.94,0.90,0.55}
\definecolor{lavenderblush}{rgb}{1.00,0.94,0.96}
\definecolor{lavender}{rgb}{0.90,0.90,0.98}
\definecolor{lawngreen}{rgb}{0.49,0.99,0.00}
\definecolor{lemonchiffon}{rgb}{1.00,0.98,0.80}
\definecolor{lightblue}{rgb}{0.68,0.85,0.90}
\definecolor{lightcoral}{rgb}{0.94,0.50,0.50}
\definecolor{lightcyan}{rgb}{0.88,1.00,1.00}
\definecolor{lightgoldenrod}{rgb}{0.93,0.87,0.51}
\definecolor{lightgoldenrod}{rgb}{0.98,0.98,0.82}
\definecolor{lightgray}{rgb}{0.83,0.83,0.83}
\definecolor{lightgreen}{rgb}{0.56,0.93,0.56}
\definecolor{lightgrey}{rgb}{0.83,0.83,0.83}
\definecolor{lightpink}{rgb}{1.00,0.71,0.76}
\definecolor{lightsalmon}{rgb}{1.00,0.63,0.48}
\definecolor{lightsea}{rgb}{0.13,0.70,0.67}
\definecolor{lightsky}{rgb}{0.53,0.81,0.98}
\definecolor{lightslate}{rgb}{0.47,0.53,0.60}
\definecolor{lightslate}{rgb}{0.47,0.53,0.60}
\definecolor{lightslate}{rgb}{0.52,0.44,1.00}
\definecolor{lightsteel}{rgb}{0.69,0.77,0.87}
\definecolor{lightyellow}{rgb}{1.00,1.00,0.88}
\definecolor{limegreen}{rgb}{0.20,0.80,0.20}
\definecolor{linen}{rgb}{0.98,0.94,0.90}
\definecolor{magenta1}{rgb}{1.00,0.00,1.00}
\definecolor{magenta2}{rgb}{0.93,0.00,0.93}
\definecolor{magenta3}{rgb}{0.80,0.00,0.80}
\definecolor{magenta4}{rgb}{0.55,0.00,0.55}
\definecolor{magenta}{rgb}{1.00,0.00,1.00}
\definecolor{maroon1}{rgb}{1.00,0.20,0.70}
\definecolor{maroon2}{rgb}{0.93,0.19,0.65}
\definecolor{maroon3}{rgb}{0.80,0.16,0.56}
\definecolor{maroon4}{rgb}{0.55,0.11,0.38}
\definecolor{maroon}{rgb}{0.69,0.19,0.38}
\definecolor{mediumaquamarine}{rgb}{0.40,0.80,0.67}
\definecolor{mediumblue}{rgb}{0.00,0.00,0.80}
\definecolor{mediumorchid}{rgb}{0.73,0.33,0.83}
\definecolor{mediumpurple}{rgb}{0.58,0.44,0.86}
\definecolor{mediumsea}{rgb}{0.24,0.70,0.44}
\definecolor{mediumslate}{rgb}{0.48,0.41,0.93}
\definecolor{mediumspring}{rgb}{0.00,0.98,0.60}
\definecolor{mediumturquoise}{rgb}{0.28,0.82,0.80}
\definecolor{mediumviolet}{rgb}{0.78,0.08,0.52}
\definecolor{midnightblue}{rgb}{0.10,0.10,0.44}
\definecolor{mintcream}{rgb}{0.96,1.00,0.98}
\definecolor{mistyrose}{rgb}{1.00,0.89,0.88}
\definecolor{moccasin}{rgb}{1.00,0.89,0.71}
\definecolor{navajowhite}{rgb}{1.00,0.87,0.68}
\definecolor{navyblue}{rgb}{0.00,0.00,0.50}
\definecolor{navy}{rgb}{0.00,0.00,0.50}
\definecolor{oldlace}{rgb}{0.99,0.96,0.90}
\definecolor{olivedrab}{rgb}{0.42,0.56,0.14}
\definecolor{orange1}{rgb}{1.00,0.65,0.00}
\definecolor{orange2}{rgb}{0.93,0.60,0.00}
\definecolor{orange3}{rgb}{0.80,0.52,0.00}
\definecolor{orange4}{rgb}{0.55,0.35,0.00}
\definecolor{orangered}{rgb}{1.00,0.27,0.00}
\definecolor{orange}{rgb}{1.00,0.65,0.00}
\definecolor{orchid1}{rgb}{1.00,0.51,0.98}
\definecolor{orchid2}{rgb}{0.93,0.48,0.91}
\definecolor{orchid3}{rgb}{0.80,0.41,0.79}
\definecolor{orchid4}{rgb}{0.55,0.28,0.54}
\definecolor{orchid}{rgb}{0.85,0.44,0.84}
\definecolor{palegoldenrod}{rgb}{0.93,0.91,0.67}
\definecolor{palegreen}{rgb}{0.60,0.98,0.60}
\definecolor{paleturquoise}{rgb}{0.69,0.93,0.93}
\definecolor{paleviolet}{rgb}{0.86,0.44,0.58}
\definecolor{papayawhip}{rgb}{1.00,0.94,0.84}
\definecolor{peachpuff}{rgb}{1.00,0.85,0.73}
\definecolor{peru}{rgb}{0.80,0.52,0.25}
\definecolor{pink1}{rgb}{1.00,0.71,0.77}
\definecolor{pink2}{rgb}{0.93,0.66,0.72}
\definecolor{pink3}{rgb}{0.80,0.57,0.62}
\definecolor{pink4}{rgb}{0.55,0.39,0.42}
\definecolor{pink}{rgb}{1.00,0.75,0.80}
\definecolor{plum1}{rgb}{1.00,0.73,1.00}
\definecolor{plum2}{rgb}{0.93,0.68,0.93}
\definecolor{plum3}{rgb}{0.80,0.59,0.80}
\definecolor{plum4}{rgb}{0.55,0.40,0.55}
\definecolor{plum}{rgb}{0.87,0.63,0.87}
\definecolor{powderblue}{rgb}{0.69,0.88,0.90}
\definecolor{purple1}{rgb}{0.61,0.19,1.00}
\definecolor{purple2}{rgb}{0.57,0.17,0.93}
\definecolor{purple3}{rgb}{0.49,0.15,0.80}
\definecolor{purple4}{rgb}{0.33,0.10,0.55}
\definecolor{purple}{rgb}{0.63,0.13,0.94}
\definecolor{red1}{rgb}{1.00,0.00,0.00}
\definecolor{red2}{rgb}{0.93,0.00,0.00}
\definecolor{red3}{rgb}{0.80,0.00,0.00}
\definecolor{red4}{rgb}{0.55,0.00,0.00}
\definecolor{red}{rgb}{1.00,0.00,0.00}
\definecolor{rosybrown}{rgb}{0.74,0.56,0.56}
\definecolor{royalblue}{rgb}{0.25,0.41,0.88}
\definecolor{saddlebrown}{rgb}{0.55,0.27,0.07}
\definecolor{salmon1}{rgb}{1.00,0.55,0.41}
\definecolor{salmon2}{rgb}{0.93,0.51,0.38}
\definecolor{salmon3}{rgb}{0.80,0.44,0.33}
\definecolor{salmon4}{rgb}{0.55,0.30,0.22}
\definecolor{salmon}{rgb}{0.98,0.50,0.45}
\definecolor{sandybrown}{rgb}{0.96,0.64,0.38}
\definecolor{seagreen}{rgb}{0.18,0.55,0.34}
\definecolor{seashell1}{rgb}{1.00,0.96,0.93}
\definecolor{seashell2}{rgb}{0.93,0.90,0.87}
\definecolor{seashell3}{rgb}{0.80,0.77,0.75}
\definecolor{seashell4}{rgb}{0.55,0.53,0.51}
\definecolor{seashell}{rgb}{1.00,0.96,0.93}
\definecolor{sienna1}{rgb}{1.00,0.51,0.28}
\definecolor{sienna2}{rgb}{0.93,0.47,0.26}
\definecolor{sienna3}{rgb}{0.80,0.41,0.22}
\definecolor{sienna4}{rgb}{0.55,0.28,0.15}
\definecolor{sienna}{rgb}{0.63,0.32,0.18}
\definecolor{skyblue}{rgb}{0.53,0.81,0.92}
\definecolor{slateblue}{rgb}{0.42,0.35,0.80}
\definecolor{slategray}{rgb}{0.44,0.50,0.56}
\definecolor{slategrey}{rgb}{0.44,0.50,0.56}
\definecolor{snow1}{rgb}{1.00,0.98,0.98}
\definecolor{snow2}{rgb}{0.93,0.91,0.91}
\definecolor{snow3}{rgb}{0.80,0.79,0.79}
\definecolor{snow4}{rgb}{0.55,0.54,0.54}
\definecolor{snow}{rgb}{1.00,0.98,0.98}
\definecolor{springgreen}{rgb}{0.00,1.00,0.50}
\definecolor{steelblue}{rgb}{0.27,0.51,0.71}
\definecolor{tan1}{rgb}{1.00,0.65,0.31}
\definecolor{tan2}{rgb}{0.93,0.60,0.29}
\definecolor{tan3}{rgb}{0.80,0.52,0.25}
\definecolor{tan4}{rgb}{0.55,0.35,0.17}
\definecolor{tan}{rgb}{0.82,0.71,0.55}
\definecolor{thistle1}{rgb}{1.00,0.88,1.00}
\definecolor{thistle2}{rgb}{0.93,0.82,0.93}
\definecolor{thistle3}{rgb}{0.80,0.71,0.80}
\definecolor{thistle4}{rgb}{0.55,0.48,0.55}
\definecolor{thistle}{rgb}{0.85,0.75,0.85}
\definecolor{tomato1}{rgb}{1.00,0.39,0.28}
\definecolor{tomato2}{rgb}{0.93,0.36,0.26}
\definecolor{tomato3}{rgb}{0.80,0.31,0.22}
\definecolor{tomato4}{rgb}{0.55,0.21,0.15}
\definecolor{tomato}{rgb}{1.00,0.39,0.28}
\definecolor{turquoise1}{rgb}{0.00,0.96,1.00}
\definecolor{turquoise2}{rgb}{0.00,0.90,0.93}
\definecolor{turquoise3}{rgb}{0.00,0.77,0.80}
\definecolor{turquoise4}{rgb}{0.00,0.53,0.55}
\definecolor{turquoise}{rgb}{0.25,0.88,0.82}
\definecolor{violetred}{rgb}{0.82,0.13,0.56}
\definecolor{violet}{rgb}{0.93,0.51,0.93}
\definecolor{wheat1}{rgb}{1.00,0.91,0.73}
\definecolor{wheat2}{rgb}{0.93,0.85,0.68}
\definecolor{wheat3}{rgb}{0.80,0.73,0.59}
\definecolor{wheat4}{rgb}{0.55,0.49,0.40}
\definecolor{wheat}{rgb}{0.96,0.87,0.70}
\definecolor{whitesmoke}{rgb}{0.96,0.96,0.96}
\definecolor{white}{rgb}{1.00,1.00,1.00}
\definecolor{yellow1}{rgb}{1.00,1.00,0.00}
\definecolor{yellow2}{rgb}{0.93,0.93,0.00}
\definecolor{yellow3}{rgb}{0.80,0.80,0.00}
\definecolor{yellow4}{rgb}{0.55,0.55,0.00}
\definecolor{yellowgreen}{rgb}{0.60,0.80,0.20}
\definecolor{yellow}{rgb}{1.00,1.00,0.00}

\def\fsu5{$\cal{F}$-$SU(5)$}

\def\m1half{$M_{1/2}$}
\def\m3half{$M_{3/2}$}
\def\m32{$M_{32}$}

\begin{document}

\title{\textbf{ \emph{ Profumo di SUSY}}: \\ Suggestive Correlations in the ATLAS and CMS High Jet Multiplicity Data}

\author{Tianjun Li}

\affiliation{State Key Laboratory of Theoretical Physics, Institute of Theoretical Physics,
Chinese Academy of Sciences, Beijing 100190, P. R. China }

\affiliation{George P. and Cynthia W. Mitchell Institute for Fundamental Physics and Astronomy,
Texas A$\&$M University, College Station, TX 77843, USA }

\author{James A. Maxin}

\affiliation{George P. and Cynthia W. Mitchell Institute for Fundamental Physics and Astronomy,
Texas A$\&$M University, College Station, TX 77843, USA }

\author{Dimitri V. Nanopoulos}

\affiliation{George P. and Cynthia W. Mitchell Institute for Fundamental Physics and Astronomy,
Texas A$\&$M University, College Station, TX 77843, USA }

\affiliation{Astroparticle Physics Group, Houston Advanced Research Center (HARC),
Mitchell Campus, Woodlands, TX 77381, USA}

\affiliation{Academy of Athens, Division of Natural Sciences,
28 Panepistimiou Avenue, Athens 10679, Greece }

\author{Joel W. Walker}

\affiliation{Department of Physics, Sam Houston State University,
Huntsville, TX 77341, USA }


\begin{abstract}
We present persistently amassing evidence that the CMS and ATLAS Collaborations may indeed be already
registering supersymmetry events at the Large Hadron Collider (LHC).  Our analysis is performed in the
context of a highly phenomenologically favorable model named No-Scale \fsu5, which represents the
unification of the ${\cal F}$-lipped $SU(5)$ Grand Unified Theory (GUT), two pairs of hypothetical
TeV-scale vector-like supersymmetric multiplets derived out of F-Theory, and the dynamically established
boundary conditions of No-Scale supergravity. We document highly suggestive correlations between the first
inverse femtobarn of observations by CMS and ATLAS, where seductive excesses in multijet events, particularly
those with nine or more jets, are unambiguously accounted for by a precision Monte-Carlo simulation of the
\fsu5 model space.  This intimate correspondence is optimized by a unified gaugino mass in the neighborhood
of $M_{1/2}$=518 GeV.  We supplement this analysis by extrapolating for the expected data profile to be
realized with five inverse femtobarns of integrated luminosity, as expected to be observed at the LHC by
the conclusion of 2011.  Significantly, we find that this luminosity may be sufficient to constitute
a SUSY discovery for the favored benchmark spectrum.  Indeed, the winds wafting our way from Geneva may
already be heavy with the delicate perfume of Supersymmetry.
\end{abstract}


\preprint{ACT-19-11, MIFPA-11-51}

\maketitle


\section{Introduction}

For some time, we have suggested rather persistently (see
Refs.~\cite{Maxin:2011hy,Li:2011hr,Li:2011gh,Li:2011rp,Li:2011fu})
that events featuring ultra-high jet multiplicities represent a rather
distinctive signal at the LHC for at least specific models of supersymmetry (SUSY).
Recently, the ATLAS~\cite{Aad:2011qa} and CMS~\cite{PAS-SUS-09-001} collaborations
have each presented data on multi-jet production which is capable of beginning to substantively
test this hypothesis.  Our personal interests are actualized in a definite way
by a model combining the ${\cal F}$-lipped $SU(5)$ Grand Unified Theory
(GUT)~\cite{Barr:1981qv,Derendinger:1983aj,Antoniadis:1987dx},
two pairs of hypothetical TeV scale vector-like SUSY multiplets with origins in
${\cal F}$-theory~\cite{Jiang:2006hf,Jiang:2009zza,Jiang:2009za,Li:2010dp,Li:2010rz},
and the dynamically established boundary conditions of No-Scale
Supergravity~\cite{Cremmer:1983bf,Ellis:1983sf, Ellis:1983ei, Ellis:1984bm, Lahanas:1986uc}.
For recent discussions, see Refs.~\cite{Benhenni:2011yt,Benhenni:2011jx}.
We have demonstrated that this model, dubbed No-Scale
\fsu5~\cite{Li:2010ws,Li:2010mi,Li:2010uu,Li:2011dw,Li:2011hr,Maxin:2011hy,Li:2011xu,Li:2011in,Li:2011gh,Li:2011rp,Li:2011fu,Li:2011xg,Li:2011ex},
possesses a distinctive collider signal of precisely the type described\footnote{
For a more complete review, the reader is directed to the appendix of Ref.~\cite{Maxin:2011hy},
and to the references therein.}.

It has been demonstrated that a majority of the bare-minimally constrained~\cite{Li:2011xu} parameter space of No-Scale $\cal{F}$-$SU(5)$,
as defined by consistency with the world average top-quark mass $m_{\rm t}$, the dynamically established boundary conditions of No-Scale supergravity,
radiative electroweak symmetry breaking, the centrally observed WMAP7 CDM relic density~\cite{Komatsu:2010fb}, and precision LEP constraints on the
lightest CP-even Higgs boson $m_{h}$~\cite{Barate:2003sz,Yao:2006px} and other light SUSY chargino and neutralino mass content,
remains viable even after careful comparison against the first 1.1~${fb}^{-1}$ of LHC data~\cite{Li:2011fu}.
Moreover, a highly favorable ``golden'' subspace~\cite{Li:2010ws,Li:2010mi,Li:2011xg} exists which may simultaneously account for the key rare process limits on
the muon anomalous magnetic moment $(g~-~2)_\mu$ and the branching ratio of the flavor-changing neutral current decays
$b \to s\gamma$ and $B_{s}^{0} \to \mu^+\mu^-$.
In addition, the isolated mass parameter responsible for the global SUSY particle mass normalization, the gaugino boundary mass $M_{1/2}$, is
dynamically determined at a secondary local minimization of the minimum of the Higgs potential $V_{\rm min}$, in a manner which is deeply
consistent with all precision measurements at the physical electroweak scale~\cite{Li:2010uu,Li:2011dw,Li:2011ex}.

In the present work, we extend prior studies which have been modeled primarily after a leading CMS
search strategy, and which have specifically deconstructed only the CMS data returns, to explicitly
include an analysis of the most recent ATLAS multi-jet data~\cite{Aad:2011qa}.
In the case of both experiments, the low statistics thus far accumulated can only place a lower
bound on the universal gaugino boundary mass $M_{1/2}$ of No-Scale \fsu5 (or equivalently on the
lightest supersymmetric particle (LSP) mass $m_{\rm LSP}$).  However,  it should be emphasized that
the models in the most favorable neighborhood of the parameter space, representing the overlap of
each of the previously mentioned constraints, do not merely passively survive the new collider
data, but moreover actively enhance the (low statistics) congruity of the results from both
detectors.  These facts should be absorbed in unison with the rather generically stable No-Scale
\fsu5 prediction of $120^{+3.5}_{-1}$~GeV for the Higgs boson mass~\cite{Li:2011xg}, which is
again consistent not only with rapidly narrowing exclusion limits from the CMS~\cite{PAS-HIG-11-022},
ATLAS~\cite{ATLAS-CONF-135,ATLAS:2011ww}, CDF and D\O~Collaborations~\cite{:2011ra}, but likewise also
with certain intriguing examples of positive low statistical excesses.

In addition to providing detailed recommendations
for which event selection modes may prove most effective in probing SUSY models with the targeted
signature, we will also attempt to project what prospective discoveries the near-term future might
hold in store, as the LHC advances both the intensity and the energy frontiers.  It will be suggested
that the regions of the No-Scale \fsu5 parameter space which best explain the current experiments may,
under a suitable signal cutting prescription, approach the gold standard signal-to-background
discovery ratio of $S/\sqrt{B+1} = 5$ at the level of about $5~{fb}^{-1}$ of integrated luminosity
at the current beam energy of $\sqrt{s} = 7~$TeV.  This luminosity is expected to be delivered to each
of the experiments by close of the calendar year.  To accomplish this, we shall emphasize the intimate statistical
fitting of the gaugino mass $M_{1/2}$=518 GeV to the ATLAS and CMS observations. We thus justify optimism that No-Scale \fsu5 is not a
lamb, silently awaiting an incremental and inevitable slaughter under steady accumulation of larger
statistics, but rather a lioness, leading the charge to explain early LHC observations.

\section{The Ultra-High Jet Multiplicity Signal of No-Scale \fsu5}

Supergravity (SUGRA) is an ubiquitous infrared limit of string theory, and forms the starting point of any
target space action, whereupon mandatory localization of the Supersymmetry (SUSY) algebra leads to general
coordinate invariance and an Einstein field theory limit.  However, only the No-Scale SUGRAs
are capable of naturally providing for SUSY breaking, while maintaining a vanishing vacuum energy
at tree level, thereby avoiding a cosmological constant which scales as a power of the Planck mass,
and facilitating the observed longevity and cosmological flatness of our Universe~\cite{Cremmer:1983bf}.
At the minimum of the null scalar potential, there exist flat directions for the vacuum expectation values
(VEVs) of the dynamic moduli specifying the geometry of the compact spacetime, thus leaving these
moduli undetermined by the classical equations of motion.  The corresponding VEVs are thus dynamically 
stabilized at loop order, by minimization of the corrected scalar potential.  The scale and structure so
imparted to the compact dimensions will directly encode the effective Planck mass in the expansive spacetime,
and also specify the coupling strengths and symmetries of all gauged interactions.  In particular, the high energy
gravitino mass $M_{3/2}$, and also the proportionally equivalent universal gaugino mass $M_{1/2}$, represent
moduli which will be established in this way.  Subsequently, all gauge mediated SUSY breaking soft-terms
will be dynamically evolved down from this boundary under the renormalization group~\cite{Ellis:1984bm,Giudice:1998bp},
establishing in large measure the low energy phenomenology, including the absence of flavor-changing neutral currents~\cite{Ellis:1981ts}. 
For some earlier attempts along these lines, see Ref.~\cite{superworld}.

Crucially, this scenario, and in particular application of the non-trivial boundary condition $B_\mu =0$ on the
on the soft SUSY breaking coupling from the bilinear Higgs mass term $\mu H_d H_u$, appears to come into its
own only when applied at a unification scale approaching the Planck mass $M_{\rm Pl}$~\cite{Ellis:2001kg,Ellis:2010jb,Li:2010ws}.
The standard scale of gauge coupling near $10^{16}$~GeV will simply not suffice.  There is an intriguing
possibility in the flipped $SU(5)$ GUT that the natural decoupling of an intermediate unification for the
$SU(2)_{\rm L} \times SU(3)_{\rm c} \Rightarrow SU(5)$ subgroup from a final unification
with the remixed hypercharge $U(1)_{\rm X}$ might be exploited to push the upper unification within 
the targeted proximity of $M_{\rm Pl}$.  With only the field content of the minimal supersymmetric standard
model (MSSM), the unification of the three gauge couplings is already sufficiently well tuned that the second
phase of running in the renormalization group equations (RGEs) is quite short.  The separation of these
scales would be equivalent to the realization of true string-scale gauge coupling unification in
free fermionic string models~\cite{Jiang:2006hf,Lopez:1995cs,Lopez:1992kg}, or the decoupling scenario in $\cal{F}$-theory
models~\cite{Jiang:2009zza,Jiang:2009za,Li:2010dp,Li:2010rz}.  It has been shown that avoiding a Landau pole for the strong
coupling constant restricts the set of vector-like multiplets which may be introduced around the TeV scale
to a pair of explicitly realized constructions~\cite{Jiang:2006hf}.  Both such modifications turn out to
enhance the (formerly negative) one-loop $\beta$-function coefficient of the strong coupling, causing it to
become precisely zero.  The flatness in the running of the strong coupling $\alpha_{\rm s}$ creates a wide
gap between the couplings $\alpha_{32} \simeq \alpha_{\rm s}$ and the much smaller $\alpha_{\rm X}$
at the intermediate unification.  This gap can only be closed by a very significant secondary running phase,
which may thus elevate the final unification scale by the necessary 2-3 orders of magnitude~\cite{Li:2010dp}.
By contrast, the same field modifications made to the standard $SU(5)$ gauge group leave the point of single unification
(which may require additional threshold corrections for consistency) close to the original GUT scale~\cite{Li:2010dp}.

The effect of these changes to the $\beta$-function coefficients on the colored gaugino, or gluino, is direct in the running down
from the high energy boundary, leading to the relation $M_3/M_{1/2} \simeq \alpha_3(M_{\rm Z})/\alpha_3(M_{32}) \simeq \mathcal{O}\,(1)$
and precipitating a conspicuously light gluino mass assignment.  Likewise, the large mass splitting expected from the heaviness
of the top quark, via its strong coupling to the Higgs, is responsible for a rather light stop squark $\widetilde{t}_1$.
The distinctively predictive $M({\widetilde{t_1}}) < M({\widetilde{g}}) < M({\widetilde{q}})$ mass hierarchy of a light stop
and gluino, both much lighter than all other squarks, is stable across the full No-Scale \fsu5 model space, but is
not precisely replicated in any constrained MSSM (CMSSM) constructions of which we are aware, and
certainly not by any of the ten standard ``Snowmass Points and Slopes'' (SPS) benchmarks~\cite{Allanach:2002nj}.
This spectrum generates a unique event topology starting from the pair production of heavy squarks
$\widetilde{q} \widetilde{\overline{q}}$, except for the light stop, in the initial hard scattering process,
with each squark likely to yield a quark-gluino pair $\widetilde{q} \rightarrow q \widetilde{g}$.  Each gluino may be expected
to produce events with a high multiplicity of virtual stops, via the (possibly off-shell) $\widetilde{g} \rightarrow \widetilde{t}$
transition, which in turn may pass through the dominant chains
$\widetilde{g} \rightarrow \widetilde{t}_{1} \overline{t} \rightarrow t \overline{t} \widetilde{\chi}_1^{0}
\rightarrow W^{+}W^{-} b \overline{b} \widetilde{\chi}_1^{0}$ and $\widetilde{g} \rightarrow \widetilde{t}_{1} \overline{t}
\rightarrow b \overline{t} \widetilde{\chi}_1^{+} \rightarrow W^{-} b \overline{b} \widetilde{\tau}_{1}^{+} \nu_{\tau}
\rightarrow W^{-} b \overline{b} \tau^{+} \nu_{\tau} \widetilde{\chi}_1^{0}$, as well as the conjugate processes
$\widetilde{g} \rightarrow \widetilde{\overline{t}}_{1} t \rightarrow t \overline{t} \widetilde{\chi}_1^{0}$ and $\widetilde{g}
\rightarrow \widetilde{\overline{t}}_{1} t \rightarrow \overline{b} t \widetilde{\chi}_1^{-}$.
The $W$ bosons will produce mostly hadronic jets and some leptons.  Employing the
{\tt MadGraph}~\cite{MGME} {\tt SDECAY} calculator, we find that a single ${\cal F}$-$SU(5)$ gluino may produce at least
4 hard jets more than 40\% of the time.  Processes similar to those described may then consistently exhibit a net product
of eight or more hard jets emergent from a single squark pair production event, passing through a single intermediate gluino pair.
When the further processes of jet fragmentation are allowed after the primary hard scattering events, this sequence
will ultimately result in a spectacular signal of ultra-high multiplicity final state jet events.
Detection by the LHC of this ultra-high jet signal could thus constitute a suggestive evocation
of the intimately linked stringy origins of $\cal{F}$-$SU(5)$, and could possibly even provide a glimpse into
the underlying structure of the fundamental string moduli.

The preferred subspace, which we shall refer to as the ``Golden Strip'', describes the model transversion of the bare-minimal constraints of Ref.~\cite{Li:2011xu} with the rare-decay processes
$b \to s \gamma$, $B_s^0 \to \mu^+ \mu^-$, and the muon anomalous magnetic moment. Recall, the bare-minimal constraints are
defined by compatibility with the world average top quark mass $m_{\rm t}$ = $173.3\pm 1.1$ GeV~\cite{:1900yx}, the prediction of a suitable candidate source of cold dark matter (CDM) relic density matching the upper and lower thresholds $0.1088 \leq \Omega_{CDM} \leq 0.1158$ set by the WMAP7
measurements~\cite{Komatsu:2010fb}, a rigid prohibition against a charged lightest supersymmetric particle (LSP), conformity with the precision LEP
constraints on the lightest CP-even Higgs boson ($m_{h} \geq 114$ GeV~\cite{Barate:2003sz,Yao:2006px}) and other light SUSY chargino, stau, and neutralino mass content, and a self-consistency specification on the dynamically evolved value of $B_\mu$ measured at the boundary scale $M_{\cal{F}}$. An uncertainty of $\pm 1$~GeV on $B_\mu = 0$ is allowed, consistent with the induced variation from fluctuation of the strong coupling within its error bounds and the expected scale of radiative electroweak (EW) corrections.

The bare-minimal constraints are further condensed by the confluence with the $b \to s \gamma$, $B_s^0 \to \mu^+ \mu^-$, and the muon anomalous magnetic moment processes to define the Golden Strip. For the experimental limits on the flavor changing neutral current process $b \rightarrow s\gamma$, we draw on the two standard deviation limits $Br(b \to s \gamma)=3.52  \pm 0.66 \times 10^{-4}$, where the theoretical and experimental errors are added in quadrature~\cite{Barberio:2007cr, Misiak:2006zs}. We likewise apply the two standard deviation boundaries $\Delta a_{\mu}=27.5 \pm 16.5 \times 10^{-10}$~\cite{Bennett:2004pv} for the anomalous magnetic moment of the muon, $(g - 2)_\mu$.  Lastly, we use the recently published upper bound of $Br(B_{s}^{0} \rightarrow \mu^{+}\mu^{-}) < 1.9 \times 10^{-8}$~\cite{Chatrchyan:2011kr} for the process $B_{s}^{0} \rightarrow \mu^+ \mu^-$. We do note nonetheless that an alternate theoretical Standard Model (SM) treatment gives ${\rm Br} (b \to s\gamma) = (2.98 \pm 0.26) \times 10^{-4}$~\cite{Becher:2006pu}, and furthermore there exist additional uncertainties attributable to the perturbative and non-perturbative QCD corrections~\cite{Misiak:2010dz}, suggesting that a relaxation of the lower bound to the vicinity of $2.75 \times 10^{-4}$ is also certainly plausible.

\section{Simulation and Selection\label{sct:selection}}

In order to make a tangible connection between the abstraction of a given theoretical construct, and the
actuality of detailed collider level observations, a sophisticated and reliable mechanism of simulation is essential.
For the Monte Carlo analysis of No-Scale \fsu5, we have adopted a suite of industry standard tools, sequentially
employing the {\tt MadGraph}~\cite{Stelzer:1994ta,MGME}, {\tt MadEvent}~\cite{Alwall:2007st}, {\tt PYTHIA}~\cite{Sjostrand:2006za}
and {\tt PGS4}~\cite{PGS4} chain.  When intending specifically to model the ATLAS or CMS detectors, we make use of the detector cards
distributed for this purpose with the {\tt PGS4} package; in the former case, we modify the card to specify an anti-$k_t$ jet
clustering algorithm, with an angular distance parameter of $0.4$.  It is necessary to seed this simulation with specific SUSY particle
mass calculations, which we obtain from {\tt MicrOMEGAs 2.1}~\cite{Belanger:2008sj}, employing a proprietary modification of
{\tt SuSpect 2.34}~\cite{Djouadi:2002ze} to run the RGEs.  For SUSY processes, we oversample the Monte Carlo and scale down to
the required luminosity, which has the effect of suppressing statistical fluctuations.  All such 2-body final state diagrams are
included in our simulation, following the procedure detailed in Ref.~\cite{Maxin:2011hy}.

We have previously emphasized the all-important role of the data selection cuts in the extraction of a rare signal from a dominant
background~\cite{Maxin:2011hy,Li:2011hr,Li:2011fu}.  For this critical final post-processing, we employ a custom script
{\tt CutLHCO}~\cite{cutlhco} (freely available for download), which also counts and compiles the associated net statistics.  In prior versions
of this software, we have focused on replicating the selection criteria common to the CMS collaboration~\cite{Maxin:2011hy,Li:2011fu}, also
presenting modifications of that basic scheme which highlight the distinctive signal of No-Scale \fsu5.  In the presently available
{\tt V1.4} release, we extend the functionality of this program to mimic selections specific to the ATLAS collaboration.
In particular, the present section algorithmically details our attempt to closely replicate the cuts described in
Ref.~\cite{Aad:2011qa}, a study of high-multiplicity jet events by the ATLAS collaboration.  We will provide the names of
selection variables to be specified in the {\tt CutLHCO} card file in typewriter font as we go, summarizing afterward
in Table~(\ref{tab:cuts}).  Before proceeding, we carefully reiterate that cuts designed for efficiency in a single given task are not
guaranteed to be suitable for any secondary purpose, and in particular, that habits established in lower jet multiplicity searches do
not necessarily carry over into the ultra-high jet multiplicity search regime

As a first step in replicating the ATLAS high jet search methodology~\cite{Aad:2011qa},
we calculate the full event missing transverse energy,
\begin{equation}
H_{\rm T}^{\rm miss} \equiv \sqrt{{\left( \sum p_{\rm T} \cos \phi \right)}^2 + {\left( \sum p_{\rm T} \sin \phi \right)}^2} \, ,
\label{EQ:HTM}
\end{equation}
defined as the magnitude of the vector sum over the momentum $p_{\rm T}$ transverse
to the beamline, where $\phi$ is the azimuthal angle about the beamline.
This sum includes all calorimeter clusters (jets, leptons and photons),
with pseudorapidity $\eta \equiv - \ln \tan(\theta/2)$ no greater than a soft cut of $({\tt CUT\_PRS} = 4.50)$.
Note that the zenith angle $\theta$ is measured from the instantaneous direction of travel of the counterclockwise
beam element, such that forward (or backward) scattering corresponds to $\eta$ equals plus (or minus) infinity,
while $\eta = 0$ is a purely transverse scattering event.
We next discard all taus, plus any light leptons which fail the classification requirements of
$p_{\rm T} \ge ({\tt CUT\_PTE} = 20)$~GeV and $|\eta| \le ({\tt CUT\_PRE} = 2.47)$ for electrons, or
$p_{\rm T} \ge ({\tt CUT\_PRM} = 10)$~GeV and $|\eta| \le ({\tt CUT\_PRM} = 2.40)$ for muons.  We note
in passing that an alternatively plausible interpretation of the ATLAS selections, which reclassifies
these discarded leptons as jet candidates, yields no net change in the final jet count per bin.
An initial filtering is then made on jets, restricting candidates to the subset satisfying 
$p_{\rm T} \ge ({\tt CUT\_PTR} = 20)$~GeV and $|\eta| \le ({\tt CUT\_PRK} = 2.80)$.
Next, jets are dismissed which lie within a cone of
$\Delta R \equiv \sqrt{{\left(\Delta \eta \right)}^2 + {\left(\Delta \phi \right)}^2} < ({\tt CUT\_RJE} = 0.2)$
around an electron.  Subsequently, any electrons within $\Delta R < ({\tt CUT\_REJ} = 0.4)$ of a surviving jet
are rejected, and likewise for any muons within the same $\Delta R < ({\tt CUT\_RMJ} = 0.4)$ perimeter.
At this point, the net scalar sum on transverse momentum
$H_{\rm T} \equiv \sum_{\rm jets} \left| {\vec{p}}_{\rm T} \right|$ may be calculated,
for all jets surviving a second round of soft cuts on $p_{\rm T} \ge ({\tt CUT\_PTS} = 40)$~GeV.
A hard cut is then made on jets with transverse momenta below a threshold established by the desired
signal region, keeping, for example, jets with $p_{\rm T} \ge ({\tt CUT\_PTC} = 55)$~GeV.
Finally, any jets within a distance of $\Delta R < ({\tt CUT\_RJJ} = 0.6)$ from another jet are pruned
in a sequenced designed to systematically minimize the number of eliminations.

Having filtered the jet and lepton candidate classifications, there are also several cuts which may
reject events as a whole.  First, any events with remaining light leptons are discarded
by zeroing out the filter on maximal electron or muon transverse momentum $({\tt CUT\_EMC} = 0.0)$.
ATLAS relies heavily in the referenced search~\cite{Aad:2011qa} on the statistic
$H_{\rm T}^{\rm miss}/\sqrt{H_{\rm T}}$, which is designed to indicate the significance of the
event missing energy relative to the expected uncertainty in the jet energy resolution.  We tabulate this
based upon the factors described previously, and subsequently execute a filter of
$H_{\rm T}^{\rm miss}/\sqrt{H_{\rm T}} \ge ({\tt CUT\_RMH} = 3.5)~{\rm GeV}^{1/2}$.
Likewise, events with less than the stipulated minimum jet multiplicity for the desired signal region, for
example $({\tt CUT\_JET} = 7)$, are also removed at this stage.  Either of these latter event cuts may be
relaxed for the purposes of sorting events will be sorted on the $H_{\rm T}^{\rm miss}/\sqrt{H_{\rm T}}$
or jet multiplicity statistics into subordinate bins for direct comparison to the ATLAS
observations.  We summarize, in the left-hand column in Table~(\ref{tab:cuts}), the parameter input which should be fed into the
{\tt CutLHCO}~\cite{cutlhco} program in order to replicate the ``ATLAS'' style cuts for the {\bf{7j55}}
signal region presently described.  The parameter option of ``UNDEF'' is a new feature available in the current
software release, for easily disabling any given selection criteria.

In the right-hand column of Table~(\ref{tab:cuts}), the prescription is given for a selection
which we shall refer to here as the ``CMS HYBRID''.  These cuts trace their origins
from our original attempt at modeling a leading ``CMS'' style search~\cite{PAS-SUS-09-001} (described carefully in
Ref.~\cite{Maxin:2011hy}), but they share important commonalities with our own suggestions for a simplified
``ULTRA'' style cut designed to emphasize an ultra-high jet multiplicity signal.  In particular, we emphasize that the
$\alpha_{\rm T}$ statistic, which is designed to suppress false missing energy diagnoses~\cite{PAS-SUS-09-001,PAS-SUS-11-003},
has been nullified. We have argued~\cite{Maxin:2011hy,Li:2011fu} that this statistic is actually actively biased against
high jet multiplicities, and that it is moreover unnecessary in this regime, given a substantial natural suppression of the
background based simply on the jet count threshold itself.  We acknowledge with much interest and applause that the
CMS collaboration has on at least one occasion provided data~\cite{PAS-SUS-11-003} which is processed without a cut on
$\alpha_{\rm T}$, {\it and} binned according to jet count, including ultra-high jet multiplicities.  The CMS HYBRID cuts
are in fact our attempt to model this approach, precisely as first demonstrated in the applicable portions of
Ref.~\cite{Li:2011fu}.  As in that reference, when binning on jet multiplicity, the $({\tt CUT\_JET} = 9)$ threshold which
we favor as a simple overarching diagnostic is relaxed for the purpose of revealing the individual per jet counts.
Although it is not yet clear how prominently these selections will be integrated into the wider search strategy of the
CMS collaboration (the role of searches which feature $\alpha_{\rm T}$ remains the principal theme of even the quoted
reference), we consider them to represent a very favorable compromise, the possible implications of which are reviewed
in Section
\ref{sct:future}.  In Section~\ref{sct:pt50}, we revisit the significance
of the larger limits on transverse momentum per jet which are preferred by both LHC collaborations than what we have
previously advocated in the ULTRA selection cuts.  Surprisingly, we unearth a dependence on the LSP mass which rescues
the higher thresholds for the currently preferred mass ranges.

\begin{table}[htbp]
        \centering
        \caption{We list the full parameter specification for our emulations of the key 
	ATLAS and CMS multi-jet event SUSY search strategies.}
                \begin{tabular}{|c|c|c|} \hline
~~Cut Name~~ & ~~ATLAS~~ & ~~CMS HYBRID~~ \\ \hline \hline
{\tt CUT\_FEM} & UNDEF & 0.9 \\ \hline
{\tt CUT\_PRC} & UNDEF & 3 \\ \hline
{\tt CUT\_PTS} & 40 & 30 \\ \hline
{\tt CUT\_PTC} & 55 & 50 \\ \hline
{\tt CUT\_PTR} & 20 & UNDEF \\ \hline
{\tt CUT\_PRK} & 2.8 & UNDEF \\ \hline
{\tt CUT\_PRS} & 4.5 & UNDEF \\ \hline
{\tt CUT\_PTE} & 20 & UNDEF \\ \hline
{\tt CUT\_PRE} & 2.47 & UNDEF \\ \hline
{\tt CUT\_PTM} & 10 & UNDEF \\ \hline
{\tt CUT\_PRM} & 2.40 & UNDEF \\ \hline
{\tt CUT\_RJE} & 0.2 & UNDEF \\ \hline
{\tt CUT\_REJ} & 0.4 & UNDEF \\ \hline
{\tt CUT\_RMJ} & 0.4 & UNDEF \\ \hline
{\tt CUT\_RJJ} & 0.6 & UNDEF \\ \hline
{\tt CUT\_JET} & 7 & 9 \\ \hline
{\tt CUT\_PTL} & UNDEF & 100 \\ \hline
{\tt CUT\_HTC} & UNDEF & 375 \\ \hline
{\tt CUT\_MET} & UNDEF & 100 \\ \hline
{\tt CUT\_PRL} & UNDEF & 2.5 \\ \hline
{\tt CUT\_ATC} & UNDEF & UNDEF \\ \hline
{\tt CUT\_RTC} & UNDEF & 1.25 \\ \hline
{\tt CUT\_PHI} & UNDEF & UNDEF \\ \hline
{\tt CUT\_PHC} & UNDEF & 25 \\ \hline
{\tt CUT\_EMC} & 0.0 & 10 \\ \hline
{\tt CUT\_RMH} & 3.5 & UNDEF \\ \hline
                \end{tabular}
\label{tab:cuts}
\end{table}

\section{Comparison to ATLAS and CMS Results}

\begin{figure*}[htp]
        \centering
        \includegraphics[width=0.45\textwidth]{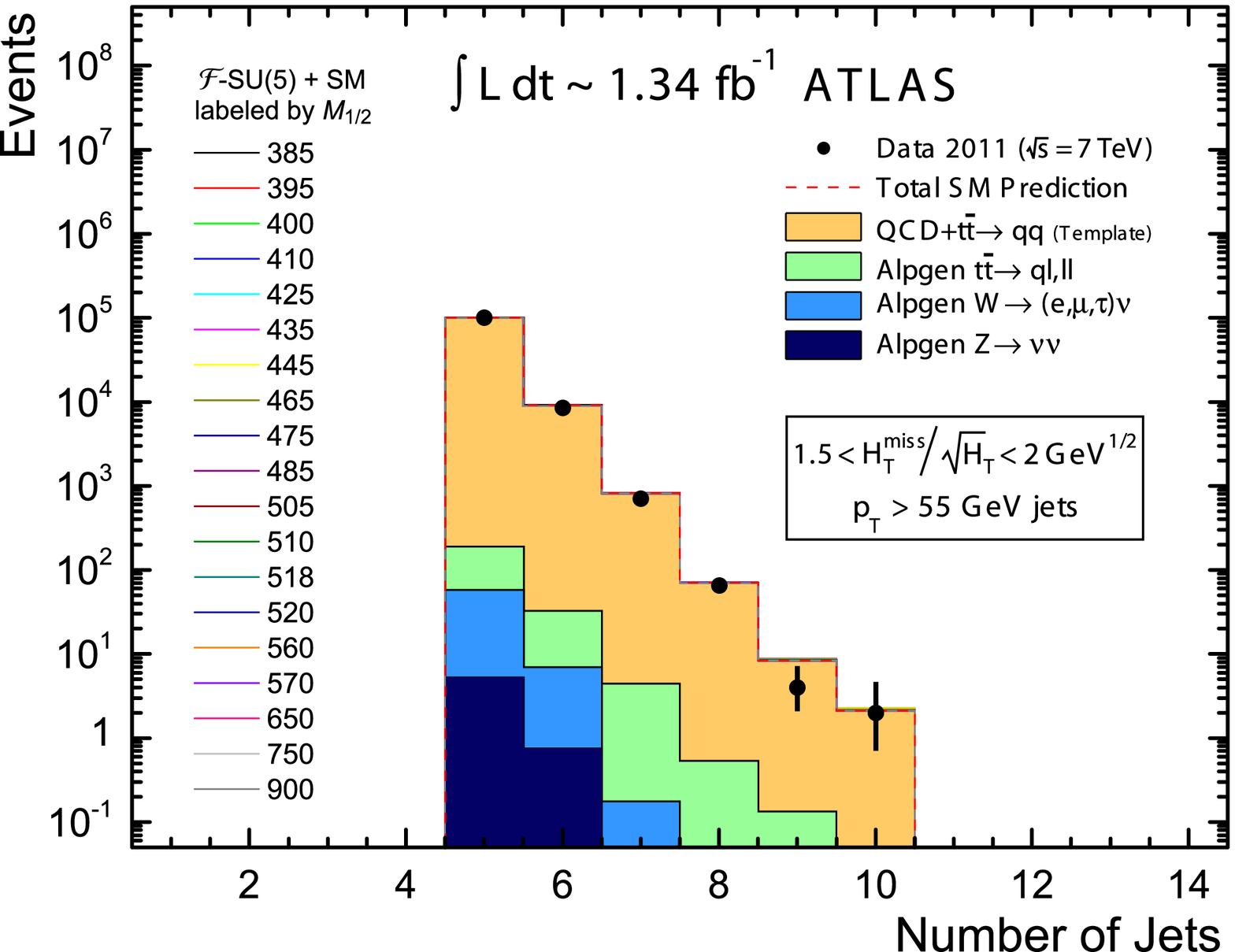}
	\hspace{0.05\textwidth}
        \includegraphics[width=0.45\textwidth]{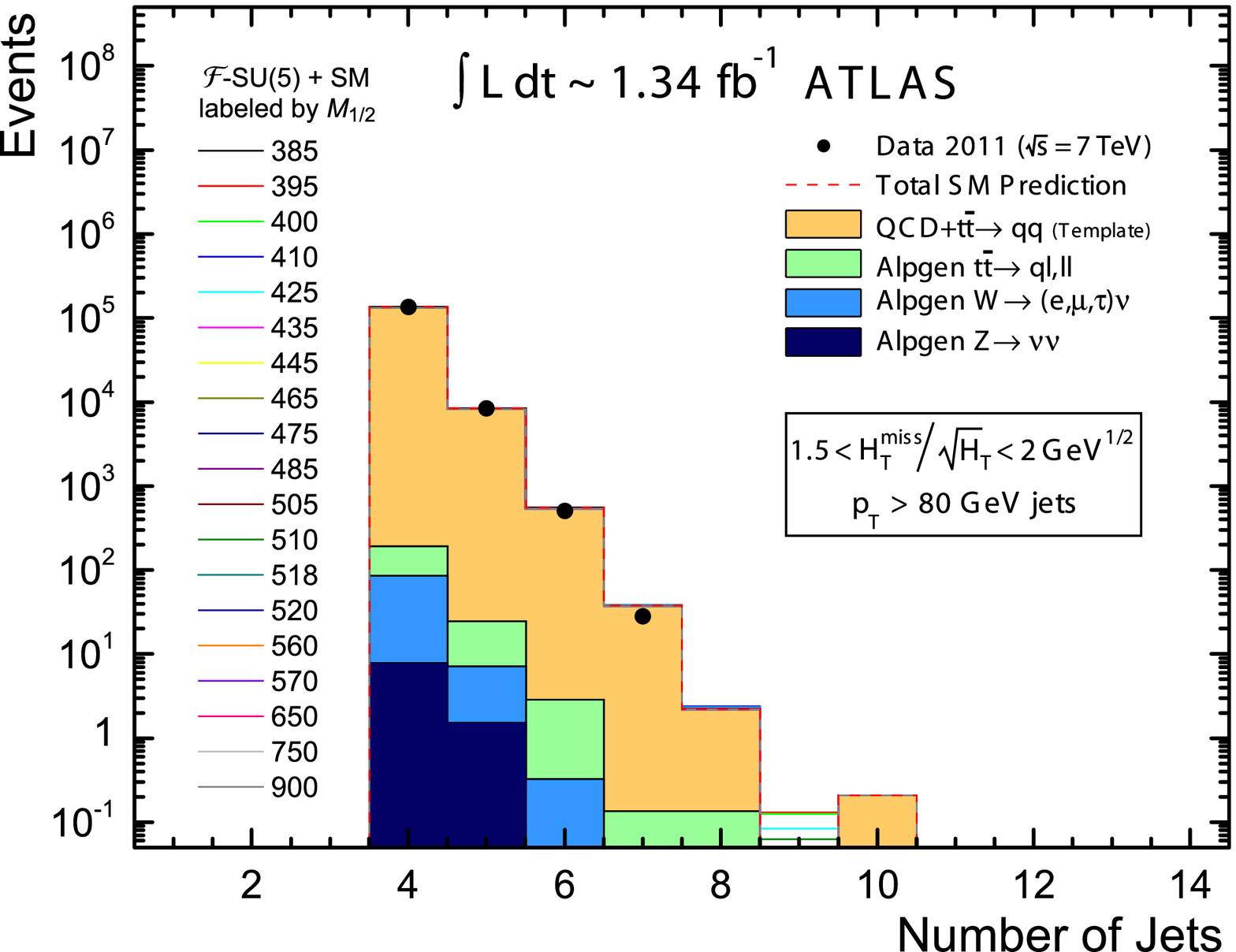} \\
	\vspace{0.05\textwidth}
        \includegraphics[width=0.45\textwidth]{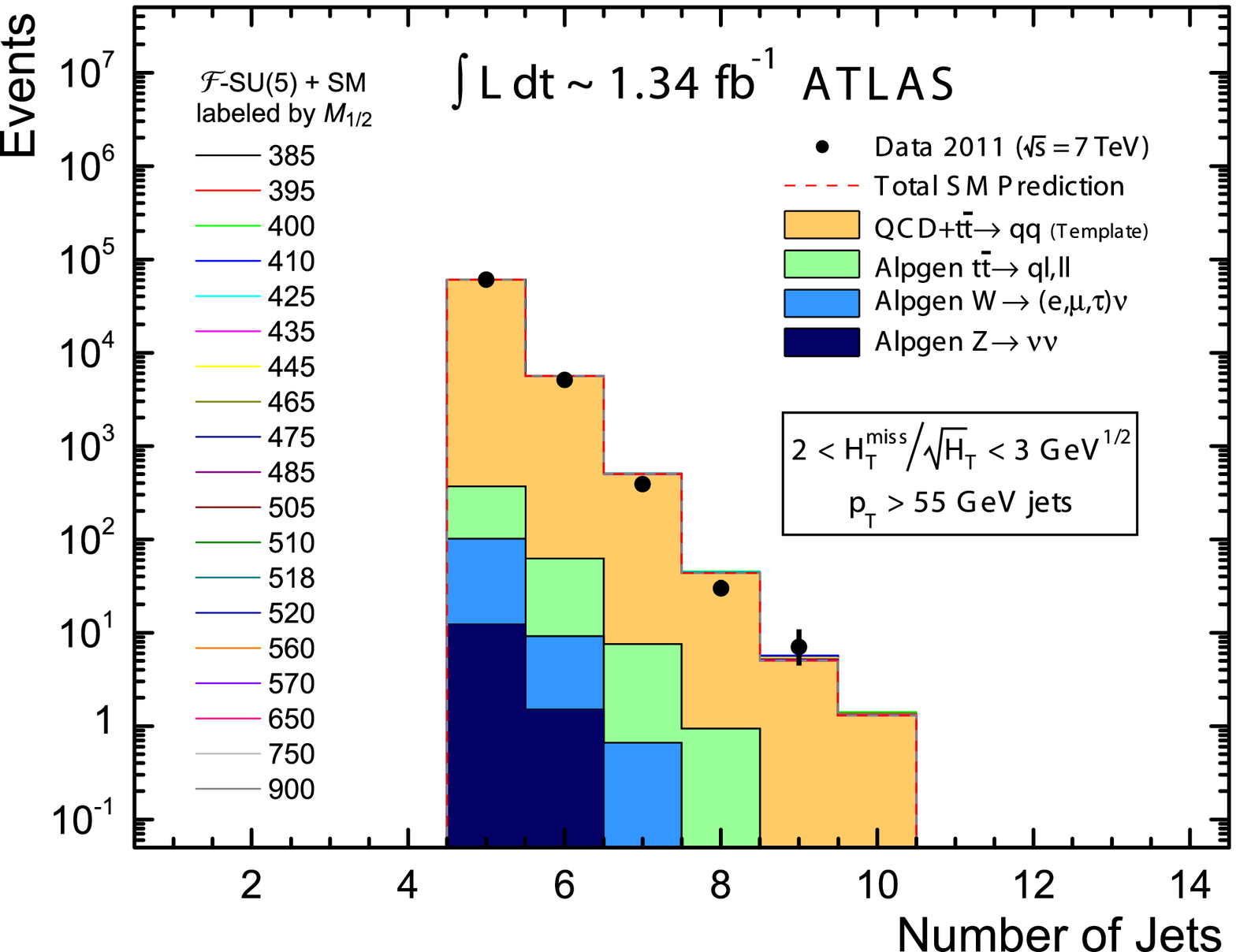}
	\hspace{0.05\textwidth}
        \includegraphics[width=0.45\textwidth]{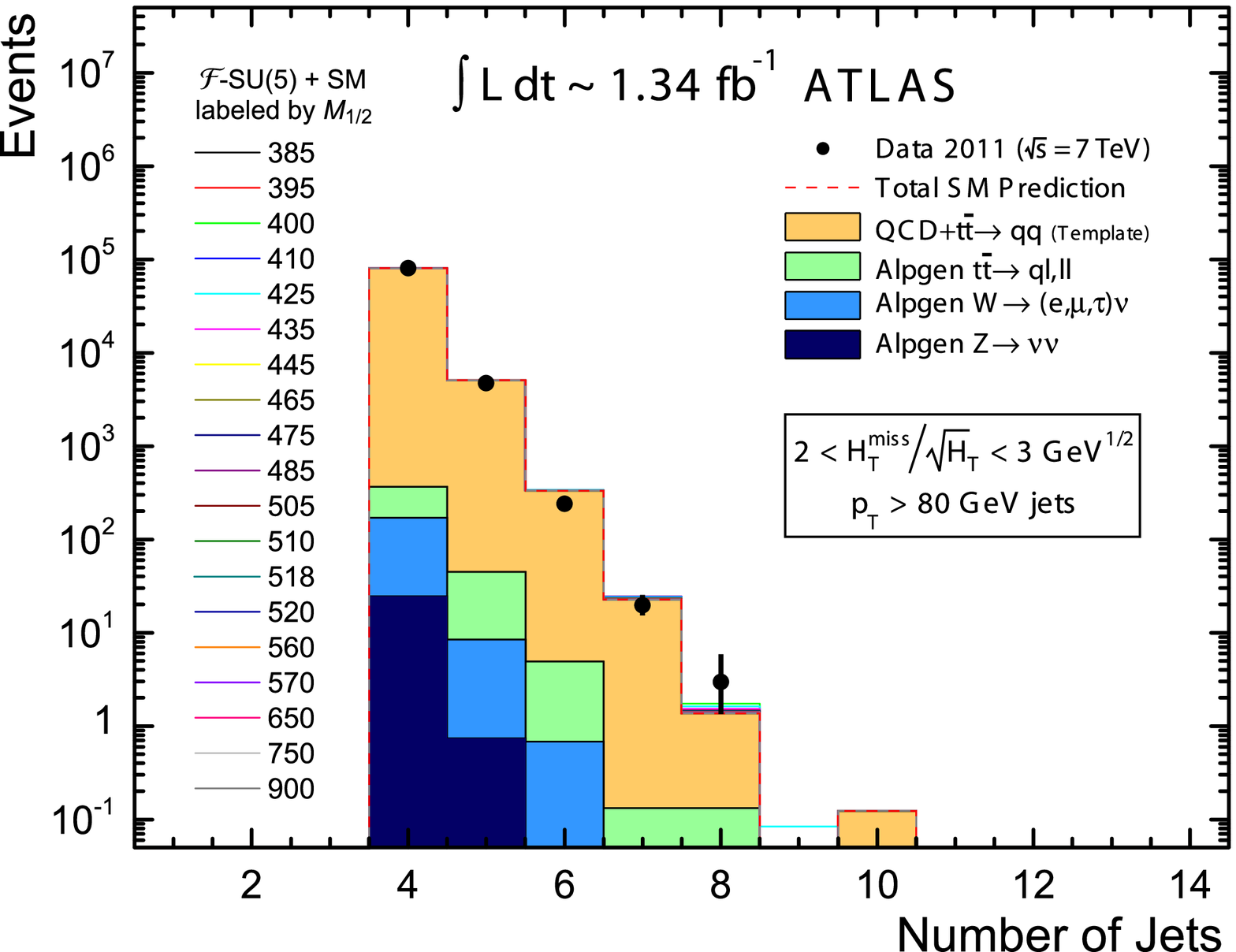} \\
	\vspace{0.05\textwidth}
        \includegraphics[width=0.45\textwidth]{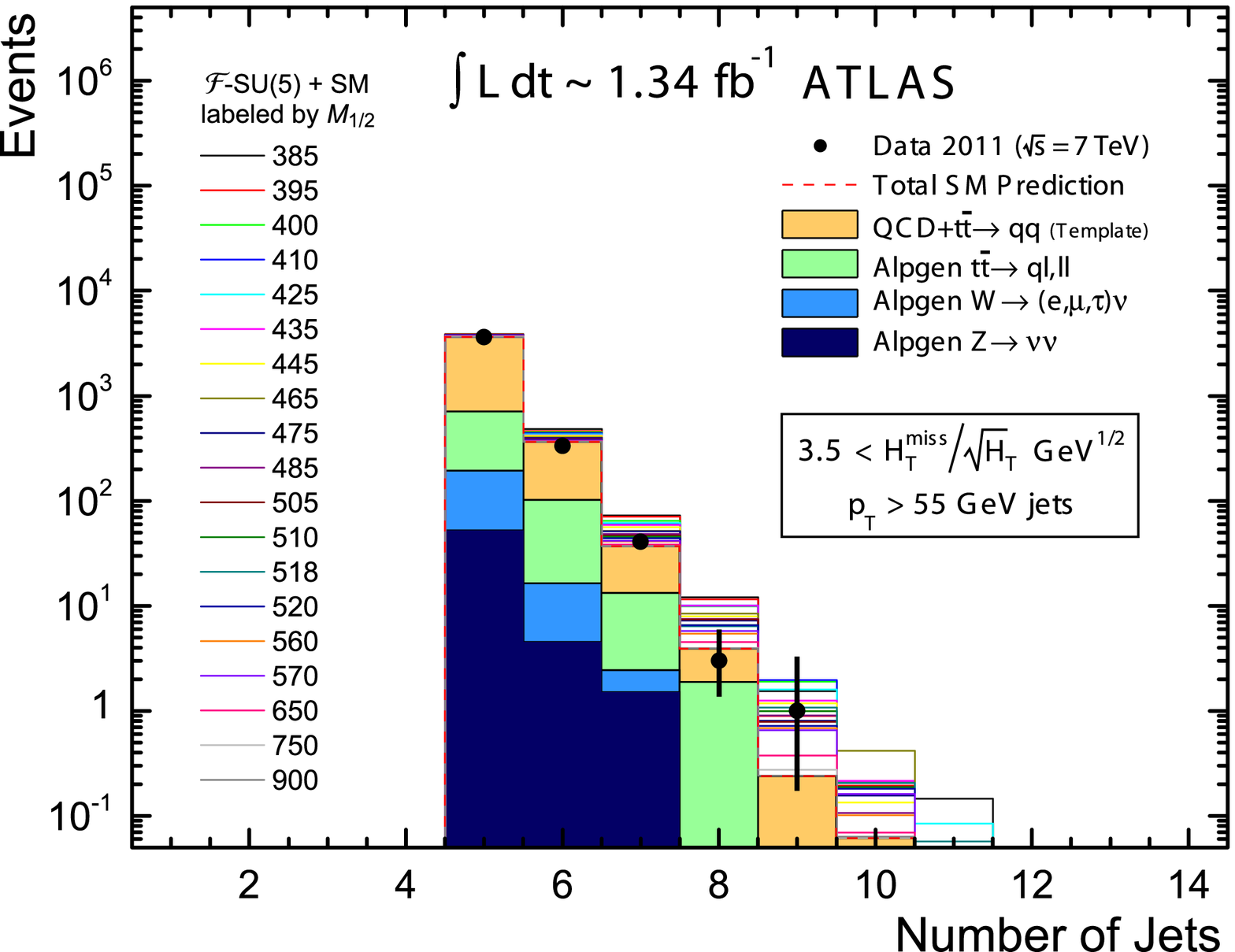}
	\hspace{0.05\textwidth}
        \includegraphics[width=0.45\textwidth]{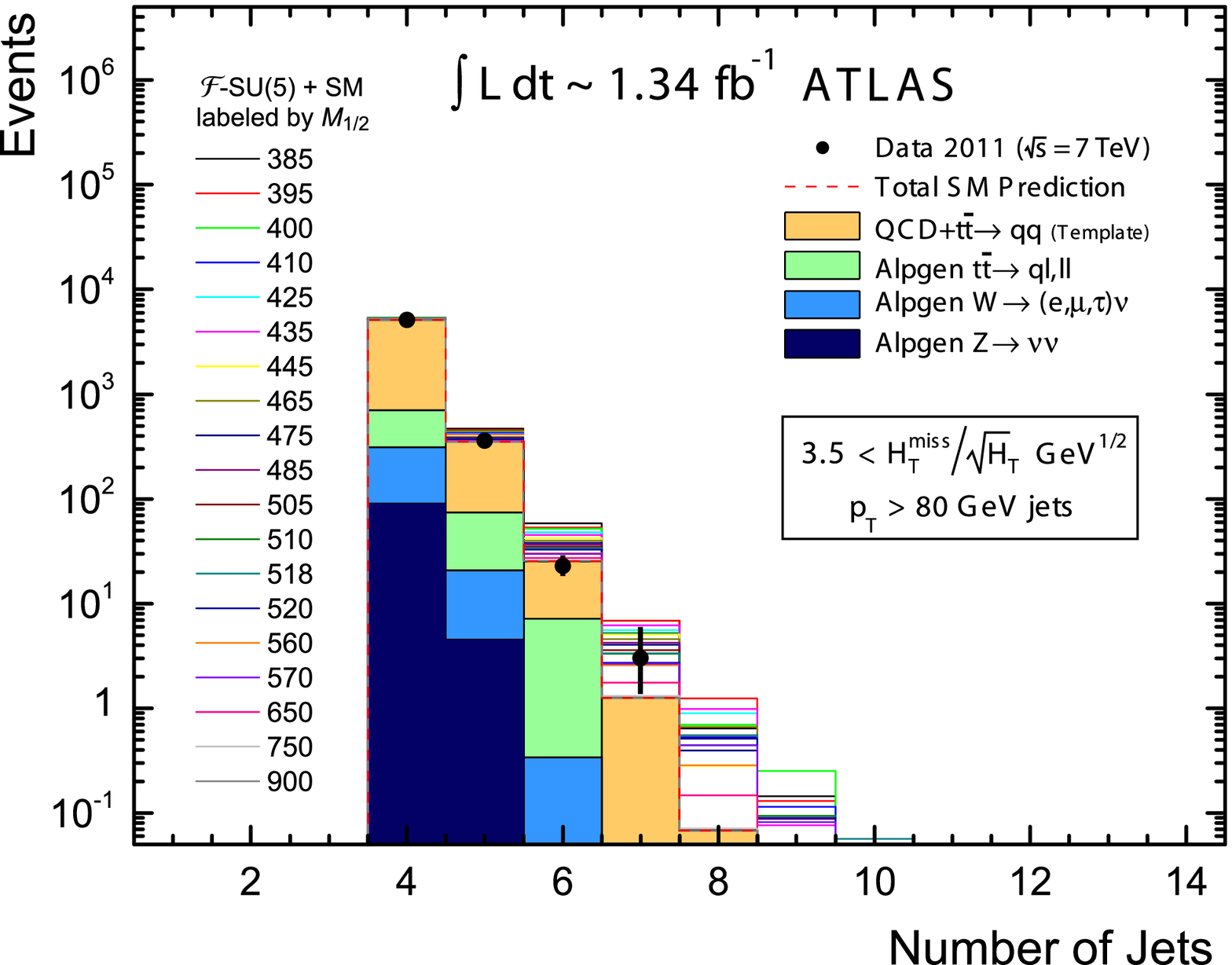}
        \caption{The ATLAS signal and background statistics for $1.34~{fb}^{-1}$ of integrated luminosity
	at $\sqrt{s} = 7$~TeV, as presented in \cite{Aad:2011qa}, are reprinted with an overlay consisting of a Monte Carlo
	collider-detector simulation of the No-Scale \fsu5 model space benchmarks of Table~\ref{tab:MCProduction}. The plot counts events per jet multiplicity. The Monte Carlo overlay consists of the \fsu5 supersymmetry signal plus the Standard Model background, thus permitting a direct visual evaluation against the ATLAS observed data points.}
        \label{fig:ATLAS_6plex}
\end{figure*}

In Figure Set~(\ref{fig:ATLAS_6plex}), we overlay our simulation of the No-Scale \fsu5
model space onto stacked-bar histograms released by the ATLAS collaboration~\cite{Aad:2011qa}
for six binning intervals, taking $H_{\rm T}^{\rm miss}/\sqrt{H_{\rm T}}$ in the ranges
$(1.5 \to 2.0)$, $(2.0 \to 3.0)$ and $(3.5 \to \inf)$, each for cuts on the transverse
momentum per jet of $55$ and $80$~GeV~\footnote{The lower two figures have been made available in
the cited arXiv preprint source package~\cite{Aad:2011qa}, but are not included in the compiled
document itself.}.  Note that the lower binning intervals of $H_{\rm T}^{\rm miss}/\sqrt{H_{\rm T}}$ are
intended chiefly as calibration spaces, with low missing energy;  consistently, our SUSY signal does
not significantly contribute in these regions.  The underlying plots feature multi-jet
observations at a center-of-mass collision energy $\sqrt{s} = 7$~TeV, and an integrated
luminosity of $1.34~{fb}^{-1}$, in conjunction with high precision Monte Carlo
and data-driven backgrounds.  Our overlays are presented in summation with the net
SM prediction, to facilitate easy comparison with the observed jet counts.  We remark
that a visual inspection suggests a superior fit to the data by the SUSY-enhanced
\fsu5 totals than by the SM in isolation.  We favor models in the range of about
$500 < M_{1/2} < 600$~GeV, corresponding to an LSP mass in the approximate range of
$100$ to $120$~GeV.  The range of substantially lighter models which appears
to be strongly excluded for overproduction is consistent with that portion of the space rejected
in our prior analysis~\cite{Li:2011fu} of the CMS multi-jet data~\cite{PAS-SUS-11-003}.  Likewise, the range of models which efficiently account for the low-statistics excesses which are actually observed also
overlaps exceedingly well with our analysis for the CMS results.

In Table~(\ref{tab:MCProduction}), we make a detailed numerical comparison of the No-Scale \fsu5
model space against the same ATLAS collaboration high jet multiplicity survey~\cite{Aad:2011qa},
using their combined reporting of four signal regions as our metric.  The statistics of merit are
the count of events with seven or more jets at a transverse momentum of at least 55~GeV ({\bf{7j55}}),
and, in a common notation, also the cases {\bf{8j55}}, {\bf{6j80}}, and {\bf{7j80}}.  In all cases,
$H_{\rm T}^{\rm miss}/\sqrt{H_{\rm T}} \ge 3.5$ is enforced, along with the other selection cuts
outlined in Section~\ref{sct:selection}.  The quantity ``$\Delta \sigma$'' represents the difference
between the experimental event count and the combined SM and \fsu5 contributions, in units of the
square root of one plus the quoted SM signal background.  A value of zero is desirable, representing precise
model and observation correspondence.  Values outside of two standard deviations are printed in strikethrough,
while values inside a single standard deviation are printed in bold.  Given unaccounted uncertainties
in the tools used to produce our estimates for \fsu5, these limits may be overly strict. 
Nevertheless, we judge the range $500 < M_{1/2} < 600$~GeV to be very satisfactory, corresponding
to an LSP mass in the approximate range of $100$ to $120$~GeV.  Restrictions are strongest from
the {\bf{6j80}} scenario, for which the reported data observation and SM background expectation are
identical.  This is in contrast to each of the other three signal regions, where at least some
production excess over the SM is observed.  

\begin{table*}[htbp]
	\caption{We compare the No-Scale \fsu5 model space against
	recent high jet multiplicity results from the ATLAS collaboration~\cite{Aad:2011qa},
	integrating over $1.34~{fb}^{-1}$ of luminosity, collected at
	a $\sqrt{s}=7$~TeV center-of-mass collision energy.
	Nineteen representative \fsu5 points are selected, each satisfying the bare-minimal
	phenomenological constraints outlined in Ref.~\cite{Li:2011xu}.
	We tabulate the number of events expected for each scenario, based on our
	own Monte Carlo simulation, under a set of selection cuts designed to mimic the
	ATLAS methodology. The ``$\Delta \sigma$'' column indicates the deviation from the observed
	data exhibited by the combined Standard Model (as reported by ATLAS) and supersymmetry \fsu5
	event expectation, in units of the square root of one plus the Standard Model event expectation.
	Positive (negative) deviations represent over-production (under-production).
	Event counts which are outside the two standard deviation boundary are indicated visually with a
	strikethrough.  Those which are inside a single standard deviation are marked in bold.
        Units of GeV are taken for the dimensionful parameters $M_{1/2}, M_{\rm V}, m_{t}~{\rm and}~m_{\rm LSP}$.
        }
	\begin{center}
	\begin{tabular}{|c|c|c|c|c||c|c||c|c||c|c||c|c|} \hline
\multicolumn{5}{|c||}{ Signal Region $\Rightarrow$} &
	\multicolumn{2}{|c||}{\bf 7j55} & \multicolumn{2}{|c||}{\bf 8j55} & \multicolumn{2}{|c||}{\bf 6j80} & \multicolumn{2}{|c|}{\bf 7j80} \\ \hline \hline
\multicolumn{5}{|c||}{ Events Observed (ATLAS) $\Rightarrow$} &
	\multicolumn{2}{|c||}{$45$} & \multicolumn{2}{|c||}{$4$} & \multicolumn{2}{|c||}{$26$} & \multicolumn{2}{|c|}{$3$} \\ \hline
\multicolumn{5}{|c||}{ Standard Model Expectation  $\Rightarrow$} &
	\multicolumn{2}{|c||}{$39\pm9$} & \multicolumn{2}{|c||}{$2.3^{+4.4}_{-0.7}$} & \multicolumn{2}{|c||}{$26\pm6$} & \multicolumn{2}{|c|}{$1.3^{+0.9}_{-0.4}$} \\ \hline
\multicolumn{5}{|c||}{ Signal Significance $S/\sqrt{B+1}$ $\Rightarrow$} &
	\multicolumn{2}{|c||}{0.95} & \multicolumn{2}{|c||}{0.94} & \multicolumn{2}{|c||}{0.00} & \multicolumn{2}{|c|}{1.12} \\ \hline \hline
$	~~~M_{1/2}~~~ $&$	~~~M_{\rm V}~~~ $&$	~\tan\beta~ $&$	~~~m_{\rm t}~~~ $&$ ~m_{\rm LSP}~ $&
	{~Events~} &$	~~~\Delta \sigma~~~ $& {~Events~} &$	~~~\Delta \sigma~~~ $& {~Events~} &$	~~~\Delta \sigma~~~ $& {~Events~} &$	~~~\Delta \sigma~~~ $ \\ \hline \hline
385 &	3575 &	19.8 &	172.5 &	75 &	\cancel{44.8} &	+6.1 &	\cancel{9.7} &	+4.4 &	\cancel{39.2} &	+7.5 &	\cancel{6.4} &	+3.1 \\ \hline
395 &	2075 &	19.7 &	172.5 &	75 &	\cancel{42.5} &	+5.8 &	\cancel{8.4} &	+3.7 &	\cancel{35.0} &	+6.7 &	\cancel{6.9} &	+3.5 \\ \hline
400 &	1450 &	19.5 &	173.7 &	75 &	\cancel{35.2} &	+4.6 &	\cancel{7.6} &	+3.2 &	\cancel{31.5} &	+6.1 &	\cancel{4.9} &	+2.1 \\ \hline
410 &	925 &	19.4 &	174.4 &	75 &	\cancel{30.9} &	+3.9 &	\cancel{7.8} &	+3.4 &	\cancel{27.2} &	+5.2 &	4.5 &		+1.8 \\ \hline
425 &	3550 &	20.4 &	172.2 &	83 &	\cancel{32.4} &	+4.2 &	\cancel{7.5} &	+3.2 &	\cancel{27.7} &	+5.3 &	\cancel{5.2} &	+2.3 \\ \hline
435 &	2000 &	20.1 &	173.1 &	83 &	\cancel{29.4} &	+3.7 &	\cancel{7.2} &	+3.1 &	\cancel{26.0} &	+5.0 &	\cancel{5.9} &	+2.8 \\ \hline
445 &	1125 &	19.9 &	174.4 &	83 &	\cancel{24.2} &	+2.9 &	5.1	 &	+1.8 &	\cancel{20.9} &	+4.0 &	4.0 &		+1.5 \\ \hline
465 &	3850 &	20.7 &	172.2 &	92 &	\cancel{20.4} &	+2.3 &	\cancel{5.6} &	+2.2 &	\cancel{18.2} &	+3.5 &	4.0 &		+1.5 \\ \hline
475 &	2400 &	20.6 &	173.1 &	92 &	18.4 &		+2.0 &	4.0 &		+1.2 &	\cancel{15.9} &	+3.1 &	\bf{3.2} &	+1.0 \\ \hline
485 &	1475 &	20.4 &	174.3 &	92 &	15.8 &		+1.5 &	4.3 &		+1.4 &	\cancel{15.0} &	+2.9 &	3.6 &		+1.2 \\ \hline
505 &	3700 &	21.0 &	172.6 &	101 &	13.9 &		+1.2 &	4.2 &		+1.4 &	\cancel{12.3} &	+2.4 &	\bf{2.8} &	+0.7 \\ \hline
510 &	2875 &	21.0 &	174.1 &	100 &	\bf{12.4} &	+1.0 &	\bf{3.5} &	+1.0 &	\cancel{10.8} &	+2.1 &	\bf{2.5} &	+0.6 \\ \hline
518 &	3050 &	21.0 &	173.3 &	102 &	\bf{10.9} &	+0.8 &	\bf{3.5} &	+1.0 &	\cancel{11.0} &	+2.1 &	\bf{2.7} &	+0.6 \\ \hline
520 &	1725 &	20.7 &	174.4 &	100 &	\bf{10.4} &	+0.7 &	\bf{3.3} &	+0.9 &	9.3 &		+1.8 &	\bf{2.0} &	+0.2 \\ \hline
560 &	1875 &	21.0 &	174.4 &	109 &	\bf{6.2} &	+0.0 &	\bf{2.0} &	+0.2 &	6.4 &		+1.2 &	\bf{1.6} &	-0.1 \\ \hline
570 &	4000 &	21.5 &	173.2 &	115 &	\bf{7.0} &	+0.2 &	\bf{2.4} &	+0.4 &	6.3 &		+1.2 &	\bf{1.9} &	+0.1 \\ \hline
650 &	4700 &	22.0 &	173.4 &	133 &	\bf{2.2} &	-0.6 &	\bf{0.7} &	-0.5 &	\bf{2.4} &	+0.5 &	\bf{0.6} &	-0.7 \\ \hline
750 &	5300 &	22.5 &	174.4 &	156 &	\bf{0.4} &	-0.9 &	\bf{0.1} &	-0.9 &	\bf{0.4} &	+0.1 &	0.1 &		-1.1 \\ \hline
900 &	6000 &	23.0 &	174.4 &	191 &	\bf{0.1} &	-0.9 &	\bf{0.0} &	-0.9 &	\bf{0.1} &	+0.0 &	0.0 &		-1.1 \\ \hline
	\end{tabular}
\end{center}
\label{tab:MCProduction}
\end{table*}

We remark that the statistical significance of the ATLAS overproduction, as gauged by the indicator of signal (observations minus background)
to background ratio $S/\sqrt{B+1}$, is quite low, never much more than one.  By contrast, as elaborated in the next section, the CMS overproduction
for a comparably luminosity is closer to $2.7$.  Given overlapping regions of compliance between the projections of No-Scale \fsu5 onto the
two data sets, we judge that the ATLAS cuts are somewhat harder than the CMS HYBRID cuts on the exemplified ultra-high multiplicity jet signal.
This is an issue which we shall document in Section~\ref{sct:future}.  Our judgment, based on this observation, is that the current CMS
multi-jet results are slightly better suited for isolating a currently favored region of the No-Scale \fsu5 model, while we simultaneously maintain basic consistency with the ATLAS results, within appropriate statistical margins.

\begin{figure*}[htp]
        \centering
        \includegraphics[width=0.49\textwidth]{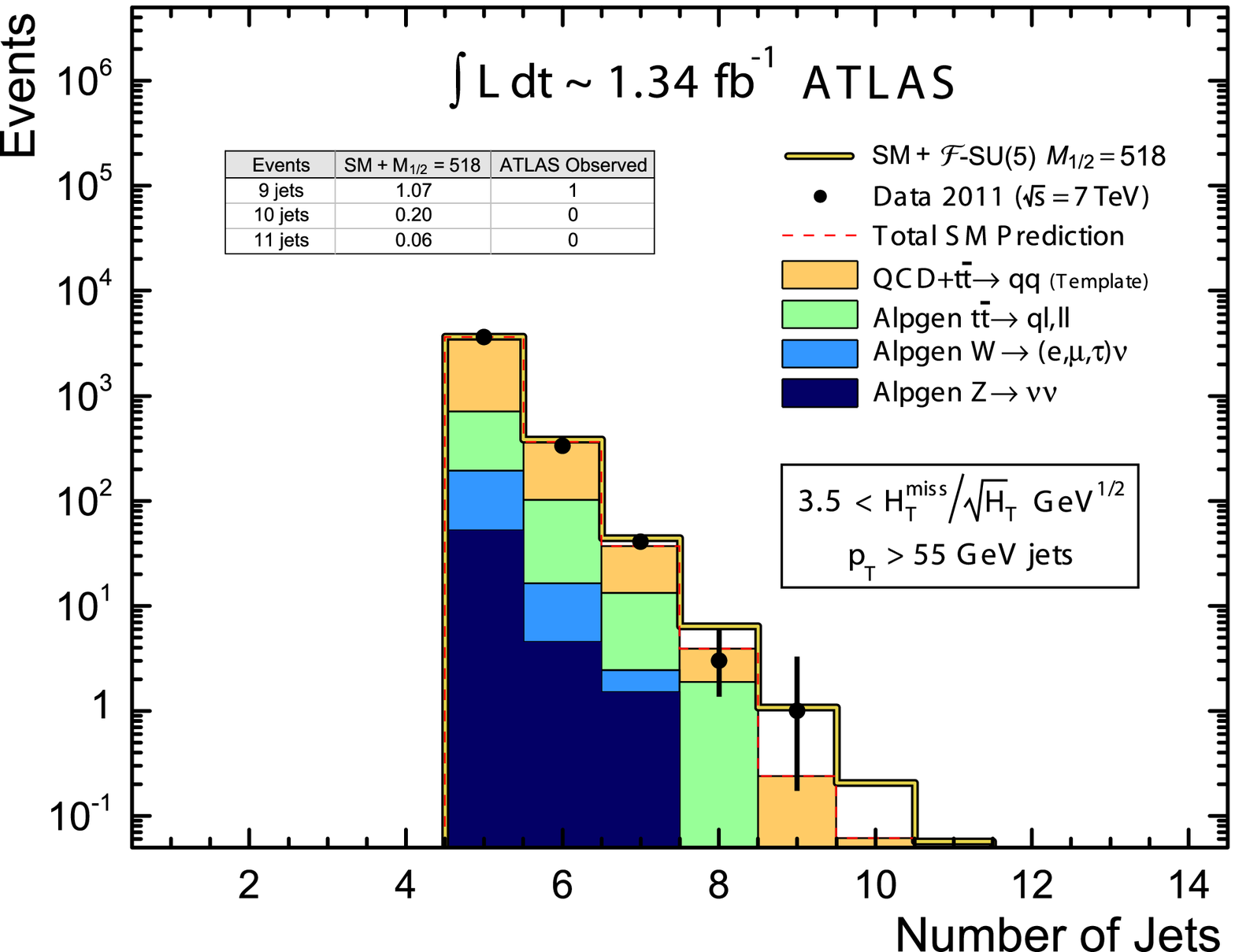}
        \includegraphics[width=0.49\textwidth]{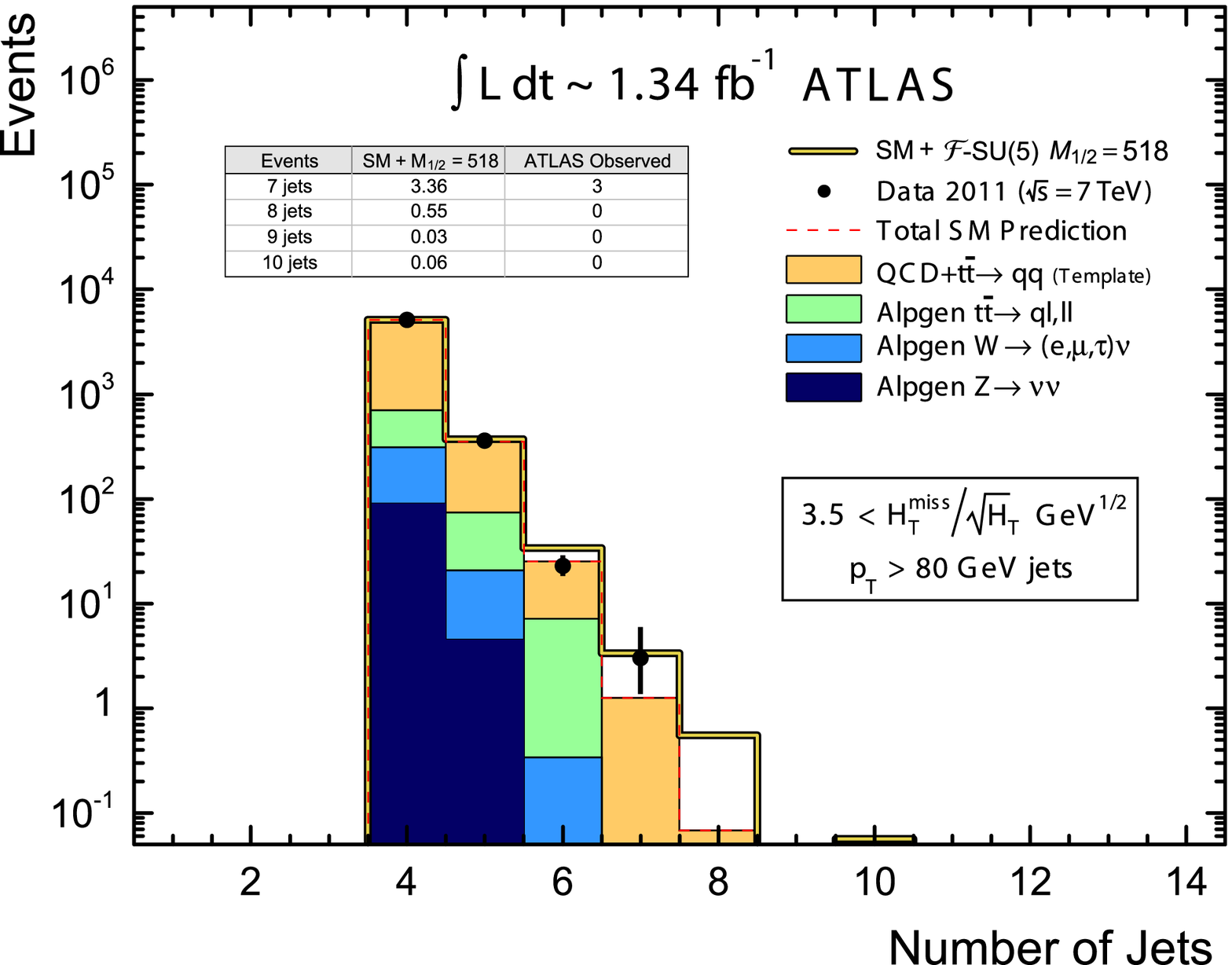}        
        \caption{The ATLAS signal and background statistics for $H_{\rm T}^{\rm miss}/\sqrt{H_{\rm T}} \ge 3.5$ for $1.34~{fb}^{-1}$ of integrated luminosity	at $\sqrt{s} = 7$~TeV, as presented in \cite{Aad:2011qa}, are reprinted with an overlay consisting of a Monte Carlo
	collider-detector simulation of the No-Scale \fsu5 model benchmark $M_{1/2}$=518 GeV for $p_T > 55$ GeV (left) and $p_T > 80$ GeV (right). The plot counts events per jet multiplicity. The Monte Carlo overlay consists of the \fsu5 supersymmetry signal plus the Standard Model background, thus permitting a direct visual evaluation against the ATLAS observed data points.}
        \label{fig:ATLAS_data}
\end{figure*}

\begin{figure*}[htp]
        \centering
        \includegraphics[width=0.45\textwidth]{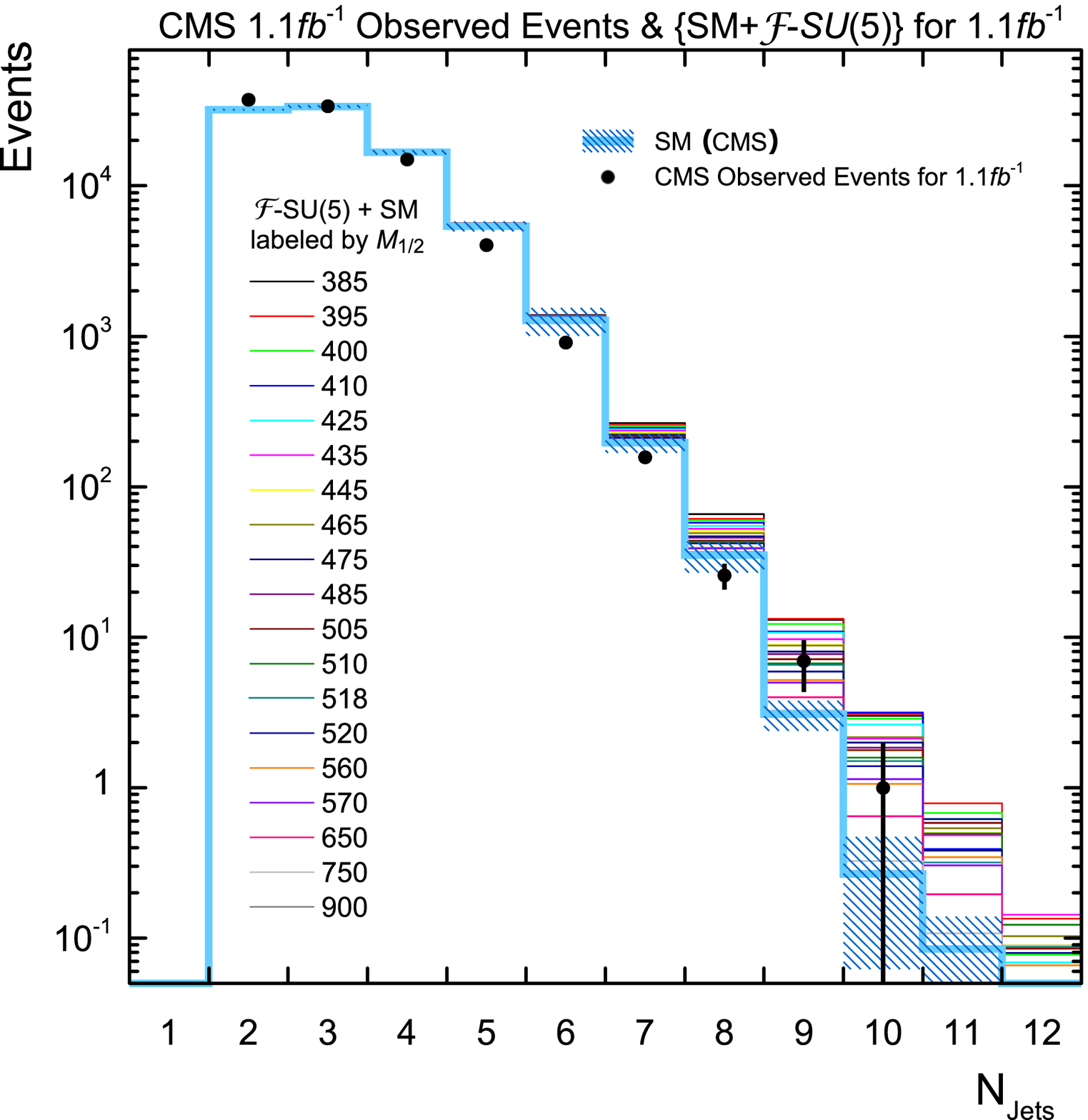}
        \includegraphics[width=0.45\textwidth]{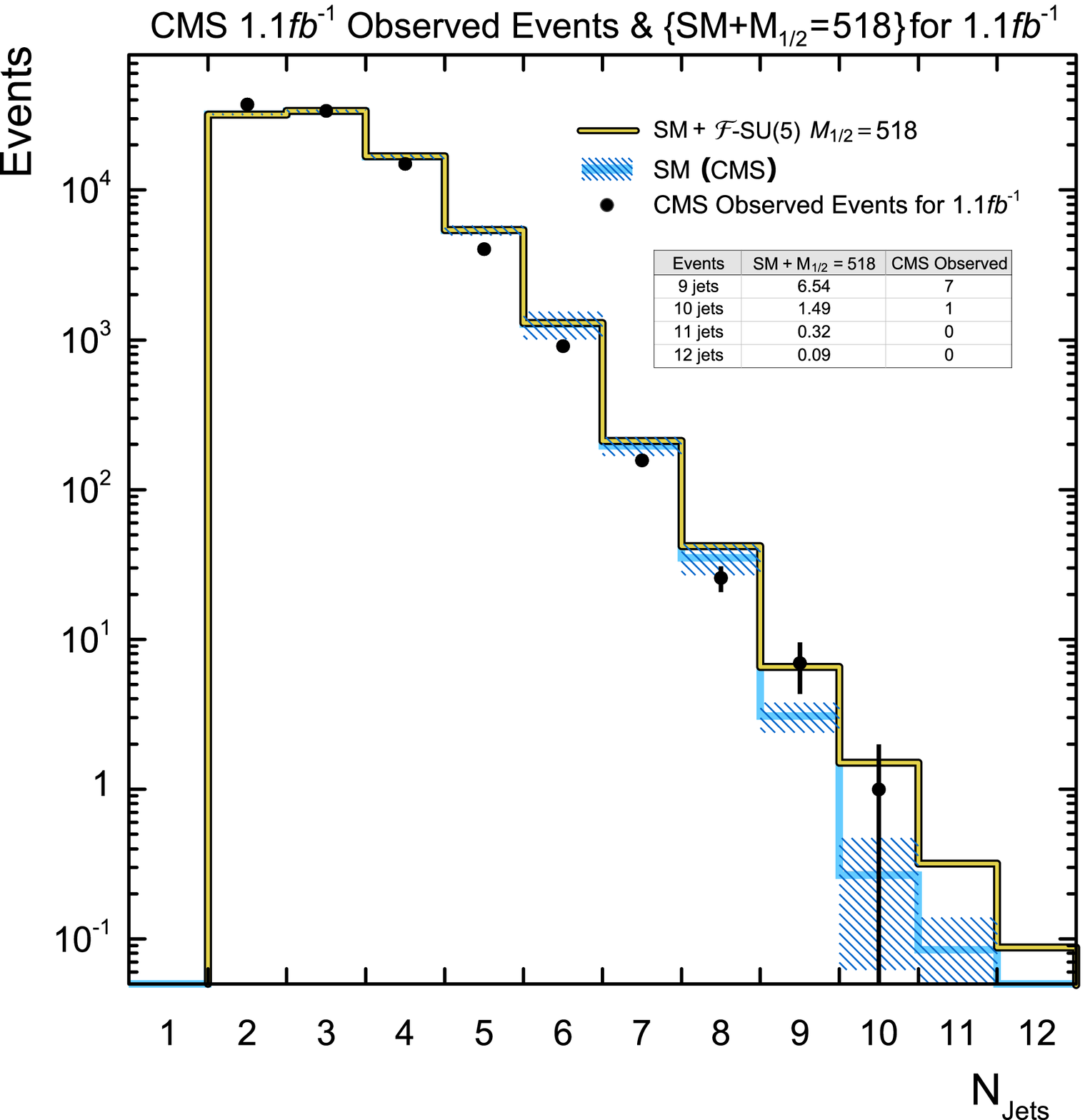}        
        \caption{The CMS signal and background statistics for $1.1~{fb}^{-1}$ of integrated luminosity
	at $\sqrt{s} = 7$~TeV, as presented in \cite{PAS-SUS-11-003}, are reprinted with an overlay consisting of a Monte Carlo
	collider-detector simulation of the No-Scale \fsu5 model space benchmarks of Table~\ref{tab:MCProduction} (left) and only the benchmark $M_{1/2}$=518 GeV (right). The plot counts events per jet multiplicity. The Monte Carlo overlay consists of the \fsu5 supersymmetry signal plus the Standard Model background, thus permitting a direct visual evaluation against the CMS observed data points.}
        \label{fig:CMS_data}
\end{figure*}

The cumulative result from application of all the experimental constraints is a narrow region of the parameter space from about $M_{1/2}$=500 GeV to 600 GeV. This Golden Strip is in accord with the region of the \fsu5 model space that maintains consistency with the CMS and ATLAS observations. The coalescence of the most favored phenomenological subspace upon the region synchronous with the CMS and ATLAS collider signals is a tantalizing development indeed. To zoom in even further in Figure Set (\ref{fig:ATLAS_6plex}) and the left plot space of Figure Set (\ref{fig:CMS_data}), we find the best fit to both the CMS and ATLAS data points within this $M_{1/2}$=500-600 GeV window is in the neighborhood of a $M_{1/2}$=518 GeV gaugino mass, as we shall elaborate next; hence we shall use this mass as the basis for our prediction at the milestone luminosity of 5 $fb^{-1}$, projected to be attained by the end of calendar year 2011. As a result, we select the model parameters $M_{1/2}$=518 GeV, $M_V$=3050 GeV, tan$\beta$=21, $m_t$=173.3 GeV for the Monte Carlo collider-detector simulation.  The full SUSY spectrum of this benchmark is provided in Table~\ref{tab:spect}.  In addition to a superior fit to the CMS and ATLAS collider signals, this point also generates a Higgs boson mass of $m_h$=$120$ GeV, consistent with the accumulated CMS, ATLAS, CDF and D\O~ statistics.

\begin{table}[htbp]
  \small
        \centering
        \caption{Spectrum (in GeV) for the favored benchmark point.
        Here, $M_{1/2}$ = 518 GeV, $\tan \beta = 21.0$, $M_{V}$ = 3050, $m_{t}$ = 173.3 GeV, $M_{Z}$ = 91.187 GeV,
        $\Omega_{\chi}$ = 0.115, $\sigma_{SI} = 1.9 \times 10^{-10}$ pb. The central prediction for
        the $p \!\rightarrow\! {(e\vert\mu)}^{\!+}\! \pi^0$ proton lifetime is around $4 \times 10^{34}$ years.
        The lightest neutralino is 99.8\% Bino.}
                \begin{tabular}{|c|c||c|c||c|c||c|c||c|c||c|c|} \hline
    $\widetilde{\chi}_{1}^{0}$&$102$&$\widetilde{\chi}_{1}^{\pm}$&$221$&$\widetilde{e}_{R}$&$196$&$\widetilde{t}_{1}$&$560$&$\widetilde{u}_{R}$&$1,027$&$m_{h}$&$120.2$\\ \hline
    $\widetilde{\chi}_{2}^{0}$&$221$&$\widetilde{\chi}_{2}^{\pm}$&$864$&$\widetilde{e}_{L}$&$556$&$\widetilde{t}_{2}$&$964$&$\widetilde{u}_{L}$&$1,116$&$m_{A,H}$&$933$\\ \hline

    $\widetilde{\chi}_{3}^{0}$&$859$&$\widetilde{\nu}_{e/\mu}$&$550$&$\widetilde{\tau}_{1}$&$111$&$\widetilde{b}_{1}$&$916$&$\widetilde{d}_{R}$&$1,066$&$m_{H^{\pm}}$&$938$\\ \hline
    $\widetilde{\chi}_{4}^{0}$&$863$&$\widetilde{\nu}_{\tau}$&$537$&$\widetilde{\tau}_{2}$&$546$&$\widetilde{b}_{2}$&$1,019$&$\widetilde{d}_{L}$&$1,119$&$\widetilde{g}$&$712$\\ \hline
                \end{tabular}
                \label{tab:spect}
\end{table}

Closely examining Figure Set (\ref{fig:ATLAS_data}) for the ATLAS observations, the highest energy binning range of $H_{\rm T}^{\rm miss}/\sqrt{H_{\rm T}} \ge 3.5$, which is the least suppressive of multijet events, shows that the $M_{1/2}$=518 GeV gaugino mass plus the Standard Model background generates the required number of observed events at nine jets for $p_T > 55$ GeV and at seven jets for $p_T > 80$ GeV. For $p_T > 55$ GeV, the $M_{1/2}$=518 GeV gaugino mass plus the Standard Model background generates 1.07 events with nine jets, when in fact only one event was observed by ATLAS. Furthermore, for $p_T > 80$ GeV, the $M_{1/2}$=518 GeV gaugino mass plus the Standard Model background produces 3.36 events with seven jets, and precisely 3 events were observed by ATLAS. Events with a higher number of jets for each of these respective cases cannot yet produce at least one event at this small luminosity of 1.34 $fb^{-1}$ in the \fsu5 plus the Standard Model background, and we find there were no events observed by ATLAS for events with more than nine jets for $p_T > 55$ GeV nor for events with more than seven jets for $p_T > 80$ GeV.

In relation to the CMS 1.1 $fb^{-1}$ data points in Figure Set (\ref{fig:CMS_data}), we see that $M_{1/2}$=518 GeV plus the Standard Model background generates 6.54 events with nine jets, and 7 events with nine jets were observed by CMS. Moreover, $M_{1/2}$=518 GeV plus the Standard Model background produces 1.49 events with ten jets, where 1 event was actually observed by CMS with ten jets. For events with greater than ten jets, the $M_{1/2}$=518 GeV gaugino mass plus the Standard Model background cannot produce at least one event at the low luminosity of 1.1 $fb^{-1}$, and in fact no events with more than ten jets were observed by CMS.

Such clear-cut correlations between the \fsu5 supersymmetry signals plus the Standard Model background with the actual observations at the LHC by CMS and ATLAS suggestively hints that \textit{the figures contained herewith may in fact already be revealing an authentic supersymmetry signal}.

\section{Revisiting the Jet Transverse Momentum Threshold\label{sct:pt50}}

\begin{figure}[htp]
        \centering
        \includegraphics[width=0.45\textwidth]{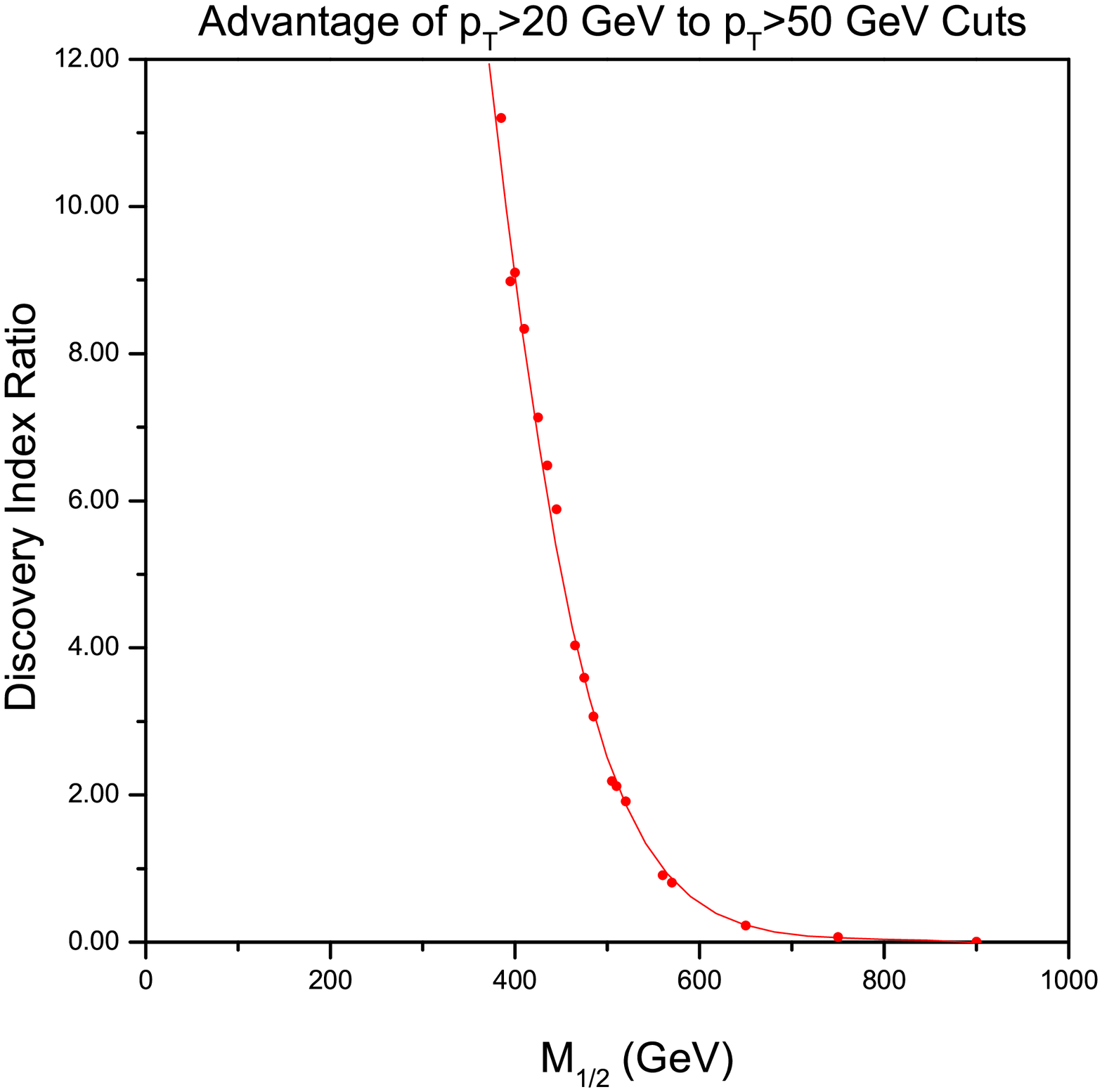}
        \caption{Dependence on $M_{1/2}$ of the improvement in the Discovery Index of a $p_T>$ 20 GeV jet cut in an event over a $p_T>$ 50 GeV cut. The $p_T>$ 20 GeV jet cut is very advantageous in the low mass regime of the model space of about $M_{1/2}<$ 500 GeV, however, the $p_T>$ 50 GeV cut on jets is superior within the larger mass region of the parameter space, at about $M_{1/2}>$ 600 GeV. The two disparate cuts appear to garner relatively equivalent results in the Golden Strip subspace of about $M_{1/2}$=500-600 GeV.}
        \label{fig:Discovery}
\end{figure}

We have previously advocated a rather low cut on the transverse momentum $p_{\rm T}$ per jet, of around 20~GeV~\cite{Maxin:2011hy,Li:2011fu},
demonstrating the superiority of that selection in contrast to more convention thresholds of around 50~GeV, under Monte Carlo simulation.
Our first analysis~\cite{Maxin:2011hy} of the preferred cutting mechanism for emphasizing the ultra-high multiplicity jet signal
exemplified by No-Scale \fsu5 was undertaken before LHC results of substantial statistical significance had been released, and well before any
suitable analyses of very high jet counts had been published.  Consequently, we focused attention at that time on the lightest portions of the model
space, which would produce the most vigorous signal.  With the advent of usable LHC statistics, we established~\cite{Li:2011fu} the first exclusion
boundaries on the No-Scale \fsu5 model space, and began to focus attention on the regions of the model featuring intermediate mass scales, which
successfully accounted for the small data excesses which began to appear.  However, we did not at this time reevaluate the consequences of variation
in the threshold on minimal jet $p_{\rm T}$, implicitly assuming that the established conventional wisdom should carry over intact.

Presently, we revisit that analysis, having now explicitly looked into scale dependencies of this cut.  We do indeed find a rather subtle dependence
on the mass of the LSP which had not previously been apparent.  The effect is summarized by Figure~(\ref{fig:Discovery}), where we plot the relative
advantage (as a ratio of required luminosities) for the discovery of No-Scale \fsu5 under $p_{\rm T} \ge 20$~GeV cuts, as opposed to
$p_{\rm T} \ge 50$~GeV, as a function of the mass parameter $M_{1/2}$.  The plot strongly reaffirms our prior results for the lighter model space,
where an advantage of a full order of magnitude applies for the low momentum cut.  However, for our presently favored mass range of
$500 < M_{1/2} < 600$~GeV, we find that the relative advantage of the weaker cuts is erased, with comparable outcomes possible for the more
conventional selection.  Moving substantially above this scale, the advantage actually turns strongly toward the higher $p_T$ cut.  This result may
come as welcome news to our experimental colleagues, who have expressed to us certain practical difficulties with implementation of the previously
advocated threshold. 

Specifically, the quantity plotted in Figure~(\ref{fig:Discovery}) is the ratio of two values of the ``discovery index'' $N$, defined in Ref.~\cite{Li:2011gh}, as
\begin{equation}
N = \frac{12.5\, B}{S^2} \times \left[ 1 + \sqrt{ 1 + {\left( \frac{2S}{5B} \right)}^2} \,\right] \, ,
\label{eq:discovery}
\end{equation}
where $S$ and $B$ are respectively the observed signal and background at some reference luminosity, conveniently chosen as $1~{fb}^{-1}$.
The value of $N$ is the relative luminosity factor by which both $S$ and $B$ should be scaled in order to
achieve a baseline value of five for the statistic of merit $S/\sqrt{B+1}$ for overall model visibility.  The backgrounds used in this
calculation are our own Monte Carlo simulation of the $t \overline{t} + {\rm jets}$ processes.  Although this sampling may be materially
incomplete, we have argued~\cite{Li:2011fu} that its sufficiency should be enhanced by use of the ratio.  The selection cuts employed are
of the CMS HYBRID variety.  We sampled also for cuts at $p_{\rm T} \ge 30$~GeV, and $p_{\rm T} \ge 40$~GeV, but find that the single demonstrated plot
successfully encapsulates the phenomenon.

The problem with the use of the $p_{\rm T} = 20$ GeV cut seems to be the large backgrounds which survive.  Of course, there is always larger event capture
overall in this case (about 15 times the net event count for the lightest models and about 5 times for the heavier models), but the statistical
significance in comparison to the background can easily turn unfavorable.  There are about 20 events which pass the
CMS HYBRID selections from $t \overline{t} + {\rm jets}$ alone, for the $p_{\rm T} \ge 20$~GeV, simulating $1~{fb}^{-1}$ of data, and that is of course
constant for all of our models.  By contrast, the elevated cut is fully efficient, leaving no background.  This is not a problem for the lighter spectra,
which can compete nicely.  But, for the heavier spectra, as the event count drops significantly below 20, the luminosity required to measure such minuscule
excesses becomes enormous.  With no background competition, the heavier cuts give much softer asymptotic behavior as the counts get low on the heavier spectra.

\section{The Once and Future LHC\label{sct:future}}

\begin{figure*}[htp]
        \centering
        \includegraphics[width=1.00\textwidth]{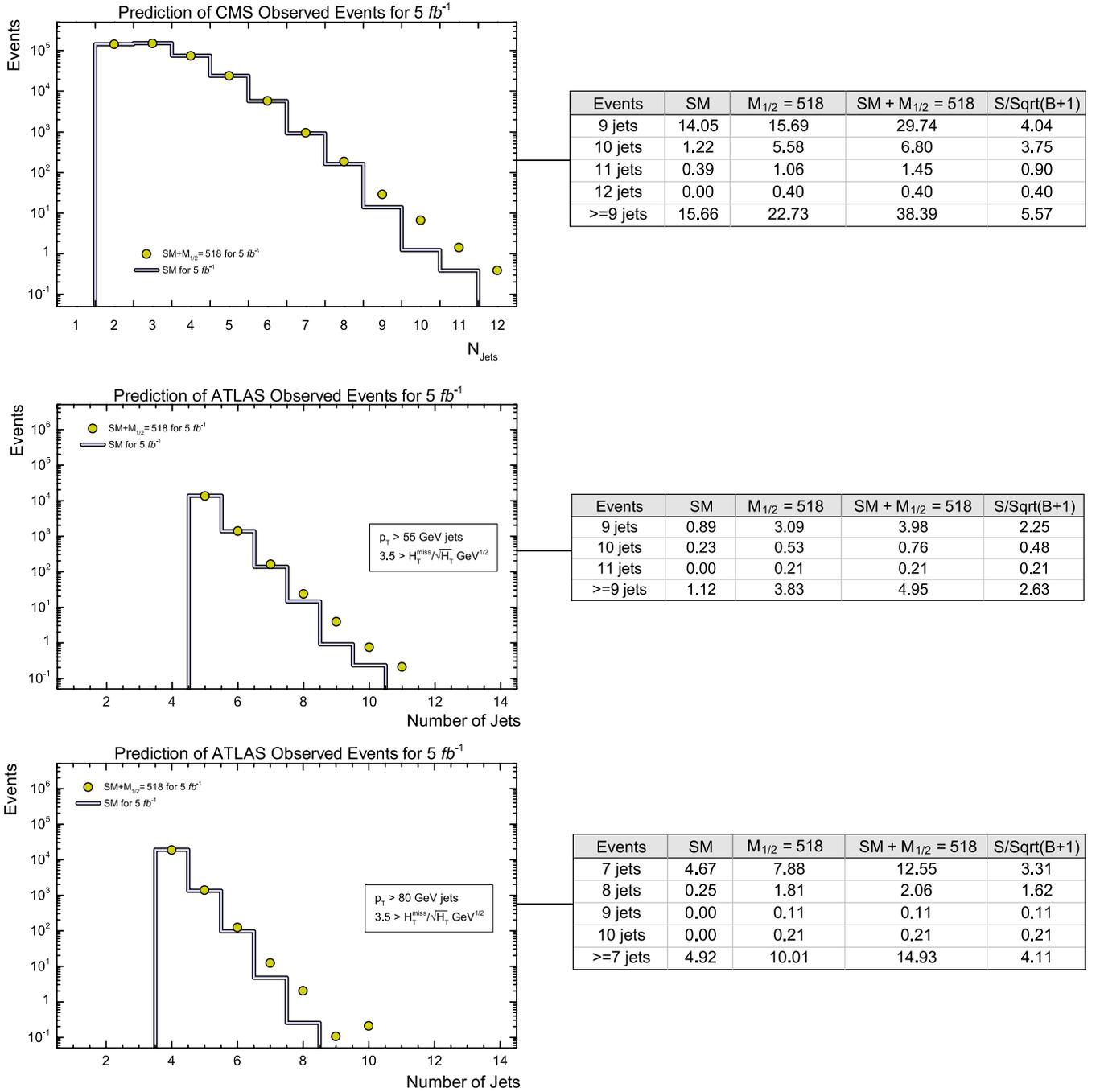}
        \caption{Predictions of the number of observed events for the CMS and ATLAS experiments for 5 $fb^{-1}$ of luminosity. The thick line represents the Standard Model expectation as determined by CMS and ATLAS, while the dots represent the sum of the Standard Model plus the simulated event count for the \fsu5 gaugino mass $M_{1/2}$=518 GeV. The tables provide the absolute predicted event count, where the final column in the tables represents the signal significance using $S/\sqrt{B+1}$. As shown, only the CMS experiment will achieve a five standard deviation signal significance for 5 $fb^{-1}$.}
        \label{fig:CMS_prediction}
\end{figure*}

\begin{table*}[h]
\renewcommand{\arraystretch}{1.0}
\begin{center}
\begin{tabular}{|c||c|c|c|c|c|c|c|c|c|c|c|c|c|c|c|c|c|c|c||c|}\hline

$M_{1/2}$&$	385$&$	395$&$	400$&$	410$&$	425$&$	435$&$	445$&$	465$&$	475$&$	485$&$	505$&$	510$&$ {\color{red} 518}$&$	520$&$	560$&$	570$&$	650$&$	750$&$	900$&$	{\rm SM} $  \\ \hline	\hline
$	{\rm 9j}  $&$	45.2$&$	46.4$&$	41.4$&$	35.3$&$	34.6$&$	29.9$&$	26.2$&$	25.9$&$	22.4$&$	21.1$&$	18.5$&$	16.2$&$ {\color{red} 15.7}$&$	12.9$&$	9.5$&$	8.6$&$	4.0$&$	0.7$&$	0.1$&$	14.1$  \\ \hline
$	{\rm 10j} $&$	12.4$&$	12.8$&$	11.9$&$	13.2$&$	10.8$&$	8.4$&$	6.8$&$	8.6$&$	7.8$&$	7.1$&$	6.9$&$	5.9$&$ {\color{red} 5.6}$&$	5.1$&$	3.6$&$	3.9$&$	1.7$&$	0.3$&$	0.1$&$	1.2$  \\ \hline
$	{\rm 11j} $&$	1.4$&$	3.2$&$	2.7$&$	1.4$&$	2.4$&$	1.8$&$	1.9$&$	2.1$&$	2.4$&$	1.9$&$	2.3$&$	1.9$&$ {\color{red} 1.1}$&$	1.4$&$	1.2$&$	1.0$&$	0.5$&$	0.1$&$	0.0$&$	0.4$  \\ \hline
$	{\rm 12j} $&$	0.4$&$	0.6$&$	0.4$&$	0.0$&$	0.3$&$	0.7$&$	0.4$&$	0.5$&$	0.2$&$	0.2$&$	0.4$&$	0.6$&$ {\color{red} 0.4}$&$	0.4$&$	0.3$&$	0.1$&$	0.2$&$	0.0$&$	0.0$&$	0.0$  \\ \hline
$	{\rm 13j} $&$	0.0$&$	0.1$&$	0.0$&$	0.1$&$	0.1$&$	0.1$&$	0.1$&$	0.2$&$	0.2$&$	0.0$&$	0.1$&$	0.1$&$ {\color{red} 0.0}$&$	0.1$&$	0.0$&$	0.2$&$	0.0$&$	0.0$&$	0.0$&$	0.0$  \\ \hline
$	{\rm 14j} $&$	0.0$&$	0.0$&$	0.0$&$	0.0$&$	0.0$&$	0.1$&$	0.0$&$	0.0$&$	0.0$&$	0.0$&$	0.0$&$	0.1$&$ {\color{red} 0.0}$&$	0.0$&$	0.0$&$	0.0$&$	0.0$&$	0.0$&$	0.0$&$	0.0$  \\ \hline	\hline
$	{\rm \ge9j} $&$	59.4$&$	63.1$&$	56.4$&$	50.0$&$	48.2$&$	41.0$&$	35.4$&$	37.3$&$	33.0$&$	30.3$&$	28.2$&$	24.8$&$ {\color{red} 22.8}$&$	19.8$&$	14.6$&$	13.7$&$	6.4$&$	1.1$&$	0.2$&$	15.7$  \\ \hline
$S/ \sqrt{B+1}$&$	\textbf{14.5}$&$	\textbf{15.4}$&$	\textbf{13.8}$&$	\textbf{12.2}$&$	\textbf{11.8}$&$	\textbf{10.0}$&$	\textbf{8.7}$&$	\textbf{9.1}$&$
	\textbf{8.1}$&$ \textbf{7.4} $&$	\textbf{6.9}$&$	\textbf{6.0}$&$	{\color{red} \textbf{5.6}}$&$ \textbf{4.8}$&$	\textbf{3.5}$&$	\textbf{3.4}$&$	\textbf{1.6}$&$	\textbf{0.27}$&$\textbf{0.05}$&-- \\ \hline
\end{tabular}
\end{center}
\caption{Predicted event counts for the CMS experiment from a Monte Carlo collider-detector simulation of the No-Scale \fsu5 model space for 5 $fb^{-1}$ of luminosity at the LHC for events with greater than or equal to nine jets. The array of $M_{1/2}$ given here consists of the benchmarks of Table~\ref{tab:MCProduction}. The number of Standard Model background events are given in the final column. The bottom row computes the signal significance at 5 $fb^{-1}$ via the ratio $S/\sqrt{B+1}$ for events with greater than or equal to nine jets.}
\label{tab:CMS_table}
\end{table*}

\begin{table*}[h]
\renewcommand{\arraystretch}{1.0}
\begin{center}
\begin{tabular}{|c||c|c|c|c|c|c|c|c|c|c|c|c|c|c|c|c|c|c|c||c|}\hline

$M_{1/2}$&$	385$&$	395$&$	400$&$	410$&$	425$&$	435$&$	445$&$	465$&$	475$&$	485$&$	505$&$	510$&$ {\color{red} 518}$&$	520$&$	560$&$	570$&$	650$&$	750$&$	900$&$	{\rm SM} $  \\ \hline	\hline
$	{\rm 9j}  $&$	4.9	$&$	2.4	$&$	6.1	$&$	6.4	$&$	5.0	$&$	3.7	$&$	3.5	$&$	2.8	$&$	2.1	$&$	2.5	$&$	2.0	$&$	2.8	$&$	{\color{red} 3.1}	$&$	1.8	$&$	1.6	$&$	1.5	$&$	0.5	$&$	0.1	$&$	0.0	$&$	0.9$	\\ \hline
$	{\rm 10j}  $&$	0.5	$&$	0.5	$&$	0.0	$&$	0.0	$&$	0.0	$&$	0.6	$&$	0.3	$&$	1.3	$&$	0.4	$&$	0.2	$&$	0.5	$&$	0.5	$&$	{\color{red} 0.5}	$&$	0.4	$&$	0.1	$&$	0.4	$&$	0.0	$&$	0.0	$&$	0.0	$&$	0.2$	\\ \hline
$	{\rm 11j}  $&$	0.5	$&$	0.0	$&$	0.0	$&$	0.0	$&$	0.3	$&$	0.0	$&$	0.0	$&$	0.0	$&$	0.0	$&$	0.2	$&$	0.0	$&$	0.0	$&$	{\color{red} 0.2}	$&$	0.1	$&$	0.1	$&$	0.0	$&$	0.0	$&$	0.0	$&$	0.0	$&$	0.0$	\\ \hline \hline
$	{\rm \ge9j}  $&$	5.9	$&$	2.9	$&$	6.1	$&$	6.4	$&$	5.3	$&$	4.3	$&$	3.8	$&$	4.1	$&$	2.5	$&$	2.8	$&$	2.5	$&$	3.3	$&$	{\color{red} 3.8}	$&$	2.3	$&$	1.9	$&$	1.9	$&$	0.6	$&$	0.1	$&$	0.0	$&$	1.1$	\\ \hline
$	S/ \sqrt{B+1}  $&$	\textbf{4.1}	$&$	\textbf{2.0}	$&$	\textbf{4.2}	$&$	\textbf{4.4}	$&$	\textbf{3.6}	$&$	\textbf{3.0}	$&$	\textbf{2.6}	$&$	\textbf{2.8}	$&$	\textbf{1.7}	$&$	\textbf{1.9}	$&$	\textbf{1.7}	$&$	\textbf{2.3}	$&$	{\color{red} \textbf{2.6}}	$&$	\textbf{1.6}	$&$	\textbf{1.3}	$&$	\textbf{1.3}	$&$	\textbf{0.4}	$&$	\textbf{0.1}	$&$	\textbf{0.0}	$& --		\\ \hline
\end{tabular}
\end{center}
\caption{Predicted event counts for the ATLAS experiment for $p_T > 55$ GeV and $H_{\rm T}^{\rm miss}/\sqrt{H_{\rm T}} \ge 3.5$ from a Monte Carlo collider-detector simulation of the No-Scale \fsu5 model space for 5 $fb^{-1}$ of luminosity at the LHC for events with greater than or equal to nine jets. The array of $M_{1/2}$ given here consists of the benchmarks of Table~\ref{tab:MCProduction}. The number of Standard Model background events are given in the final column. The bottom row computes the signal significance at 5 $fb^{-1}$ via the ratio $S/\sqrt{B+1}$ for events with greater than or equal to nine jets.}
\label{tab:ATLAS_table1}
\end{table*}

\begin{table*}[h]
\renewcommand{\arraystretch}{1.0}
\begin{center}
\begin{tabular}{|c||c|c|c|c|c|c|c|c|c|c|c|c|c|c|c|c|c|c|c||c|}\hline

$M_{1/2}$&$	385$&$	395$&$	400$&$	410$&$	425$&$	435$&$	445$&$	465$&$	475$&$	485$&$	505$&$	510$&$ {\color{red} 518}$&$	520$&$	560$&$	570$&$	650$&$	750$&$	900$&$	{\rm SM} $  \\ \hline	\hline
$	{\rm 7j}  $&$	21.1	$&$	21.0	$&$	15.1	$&$	14.6	$&$	16.2	$&$	18.4	$&$	14.3	$&$	12.6	$&$	10.4	$&$	11.1	$&$	8.8	$&$	7.7	$&$	{\color{red} 7.9}	$&$	5.6	$&$	4.9	$&$	5.3	$&$	1.9	$&$	0.3	$&$	0.0	$&$	4.7	$	\\ \hline
$	{\rm 8j}  $&$	2.2	$&$	4.4	$&$	2.4	$&$	1.7	$&$	3.1	$&$	3.5	$&$	0.8	$&$	2.2	$&$	1.2	$&$	1.8	$&$	1.7	$&$	1.4	$&$	{\color{red} 1.8}	$&$	1.7	$&$	0.8	$&$	1.4	$&$	0.3	$&$	0.0	$&$	0.0	$&$	0.3	$	\\ \hline
$	{\rm 9j}  $&$	0.5	$&$	0.5	$&$	0.9	$&$	0.4	$&$	0.0	$&$	0.3	$&$	0.0	$&$	0.0	$&$	0.2	$&$	0.3	$&$	0.0	$&$	0.4	$&$	{\color{red} 0.1}	$&$	0.3	$&$	0.1	$&$	0.3	$&$	0.0	$&$	0.0	$&$	0.0	$&$	0.0	$	\\ \hline
$	{\rm 10j}  $&$	0.0	$&$	0.0	$&$	0.0	$&$	0.0	$&$	0.0	$&$	0.0	$&$	0.0	$&$	0.0	$&$	0.2	$&$	0.2	$&$	0.0	$&$	0.0	$&$	{\color{red} 0.2}	$&$	0.0	$&$	0.0	$&$	0.0	$&$	0.0	$&$	0.0	$&$	0.0	$&$	0.0	$	\\ \hline \hline
$	{\rm \ge7j}  $&$	23.8	$&$	25.9	$&$	18.4	$&$	16.7	$&$	19.4	$&$	22.1	$&$	15.1	$&$	14.8	$&$	11.9	$&$	13.4	$&$	10.5	$&$	9.5	$&$	{\color{red} 10.0}	$&$	7.6	$&$	5.9	$&$	7.0	$&$	2.2	$&$	0.3	$&$	0.0	$&$	4.9	$	\\ \hline
$	S/ \sqrt{B+1}  $&$	\textbf{9.8}	$&$	\textbf{10.6}	$&$	\textbf{7.5}$&$	\textbf{6.9}	$&$	\textbf{8.0}	$&$	\textbf{9.1}	$&$	\textbf{6.2}	$&$	\textbf{6.1}	$&$	\textbf{4.9}	$&$	\textbf{5.5}	$&$	\textbf{4.3}	$&$	\textbf{3.9}	$&$	{\color{red} \textbf{4.1}}	$&$	\textbf{3.1}	$&$	\textbf{2.4}	$&$	\textbf{2.9}	$&$	\textbf{0.9}	$&$	\textbf{0.1}	$&$	\textbf{0.0}	$& --	\\ \hline
\end{tabular}
\end{center}
\caption{Predicted event counts for the ATLAS experiment for $p_T > 80$ GeV and $H_{\rm T}^{\rm miss}/\sqrt{H_{\rm T}} \ge 3.5$ from a Monte Carlo collider-detector simulation of the No-Scale \fsu5 model space for 5 $fb^{-1}$ of luminosity at the LHC for events with greater than or equal to seven jets. The array of $M_{1/2}$ given here consists of the benchmarks of Table~\ref{tab:MCProduction}. The number of Standard Model background events are given in the final column. The bottom row computes the signal significance at 5 $fb^{-1}$ via the ratio $S/\sqrt{B+1}$ for events with greater than or equal to seven jets.}
\label{tab:ATLAS_table2}
\end{table*}

There are two frontiers that the LHC is expected to continue probing with greater efficacy in the coming months and years.
The first is the intensity frontier, marked by the incredibly rapid escalation from the initial reports of early 2011,
featuring a few dozen picobarns of data~\cite{Khachatryan:2011tk, daCosta:2011hh, daCosta:2011qk}, to the current
Fall and end of Summer~\cite{Aad:2011qa,PAS-SUS-09-001} publications across the femtobarn threshold, and the anticipated
round of papers for the first $5~{fb}^{-1}$ expected to be delivered to each detector by the close of 2011.
It is with this luminosity in mind that we have projected our expectations for future findings at the LHC,
assuming $M_{1/2} = 518 GeV$ in Figure~(\ref{fig:CMS_prediction}) and tabulating more
general results in Tables~(\ref{tab:CMS_table})-(\ref{tab:ATLAS_table2}).
Critically, as demonstrated in Tables~(\ref{tab:CMS_table})-(\ref{tab:ATLAS_table2}), the statistical significance of the projected
results for our favored $500 < M_{1/2} < 600$~GeV model space may surpass the benchmark value of five for the lighter
of the favored models for the CMS experiment, and is expected somewhat more generically to be larger than three.
We contrast that the cuts presently exemplified by the ATLAS collaboration appear to be slightly harder on the ultra-high
jet multiplicity content, and a discover at $5~{fb}^{-1}$ appears unlikely for the models within the viable mass range.

The second very interesting possibility is an upgrade on the energy frontier side.  Of course, the LHC was originally
designed to operate at 14~TeV, but the disastrous early failure of certain poor connections in the soldering has forced
a fairly cautious institutional stance to take hold, until the joints in question may all be retrofitted with protective bypass
systems and thoroughly tested.  However, serious debates are ongoing within the LHC leadership about the risk-to-benefit
tradeoffs which would accompany a transition from $\sqrt{s} = 7$~TeV to 10 TeV.  There is no doubt that such a move would be highly
favorable for our model, assuming that it could be made safely.  We have previously made explicit comparisons between the current
7 TeV beam and speculative upgrades to 8,10,12 and 14 TeV~\cite{Li:2011gh, Li:2011rp}.  We briefly summarize our most recent analysis,
for a representative model in the favored $500 < M_{1/2} < 600$~GeV range, giving advantages in the discovery index of Eq.~(\ref{eq:discovery}),
relative to the 7 TeV beam, applying the CMS HYBRID selection cuts.  An 8 TeV beam is expected to be around twice as productive, a 10 TeV beam is
expected to be around 10 times as productive, a 12 TeV beam is expected to be around 25 times as productive, and a 14 TeV beam is expected to be
around 40 times as productive.  Again, these numbers represent the ratio of integrated luminosity which we project to be required for SUSY
discovery at the various beam energies.  In other words, they may be correctly interpreted as time efficiencies, and we suspect that one year of
running at 10 TeV might be equivalent to a decade of running at 7 TeV, with respect to the visibility of our particular model,
and given selection cuts tuned for the 9+ jet signal.

\section{Conclusions}

If we could somehow peer into the future, the imagination can wonder what events we may glimpse. In some small manner, this is exactly what we attempt to accomplish here. History has taught us that significant advancements in physics begin with a theoretical prediction, subsequently necessitating an experimental verification that cements the new theory as dogma within science. We thus embarked upon a bold journey to lay all our cards on the line, so to speak, and introduce a high-precision forecast of future collider signals to be observed by the end of this calendar year at the LHC, inspired by the convergence of the ${\cal F}$-$SU(5)$ supersymmetry event profile upon the recently published observations by the CMS and ATLAS Collaborations.

Constructed from the tripodal foundation of the ${\cal F}$-lipped $SU(5)$ GUT, extra TeV-scale vector-like multiplets with origins in F-Theory, and the dynamics of No-Scale supergravity, the ${\cal F}$-$SU(5)$ framework has demonstrated a rare consistency between parameters determined dynamically (top-down approach) and experimentally constrained parameters (bottom-up approach). Our numerous explorations of the ${\cal F}$-$SU(5)$ model have revealed deep fundamental correlations, motivating our thrust to statistically measure its collider signal spectrum in relation to currently ongoing tests of supersymmetry by CMS and ATLAS at the LHC.

A traditional phrase, originally borrowed from America's pastime of baseball, has long been ``three strikes and you're out'', though here we coin a whimsical twist fitting to our continuing challenge, and that is ``three strikes and you're in''. We have demonstrated that the ${\cal F}$-$SU(5)$ is consistent with all three major tests of supersymmetry and particle physics presently underway at the colliders. First, in~\cite{Li:2011fu} and extended upon here, we presented an in-depth analysis of the CMS 1.1 $fb^{-1}$ collider signals and illustrated through a precision Monte Carlo analysis that the ${\cal F}$-$SU(5)$ readily explains compelling data excesses in multijet events possessing more than nine jets. As such, we established a lower boundary on the gaugino mass of $M_{1/2}$=485 GeV, with the most engaging alignment between the actual observations and the supersymmetry simulations residing in the $M_{1/2}$=500-600 GeV ``golden'' subspace, the ${\cal F}$-$SU(5)$'s most phenomenologically desirable region, in the neighborhood of $M_{1/2}$=518 GeV.

Second, we submitted a previous prediction in~\cite{Li:2011xg} that the Higgs boson mass within the generic ${\cal F}$-$SU(5)$ parameter space persists at a fairly constrained $m_h$=$120^{+3.5}_{-1}$ GeV. The rapidly confining exclusion boundaries determined by the CMS, ATLAS, CDF and D\O~Collaborations are consistent with a 120 GeV ${\cal F}$-$SU(5)$ Higgs boson mass.

Third, and as expounded upon in this work in great detail, we showed that the ${\cal F}$-$SU(5)$ supersymmetry signals are not just consistent with, but exhibit a very strong correlation to the recently published ATLAS observations at the LHC. Duplicating the ATLAS jet cutting strategy and investigating the ATLAS preferred six binning intervals for the statistic $H_{\rm T}^{\rm miss}/\sqrt{H_{\rm T}}$, we presented sharp correlations between the ${\cal F}$-$SU(5)$ Monte Carlo and the ATLAS 1.34 $fb^{-1}$ of observed data, supplementing the unequivocal correspondence witnessed in the CMS search methodology. In similar accordance with the conclusions of our CMS analysis, we uncover here that the golden subspace of $M_{1/2}$=500-600 GeV is generally consistent, within two standard deviations, with the ATLAS reported collider signals.

The congruent results amongst two independently operated and managed detectors at the LHC, in parallel with the conflux of the ${\cal F}$-$SU(5)$ 120 GeV Higgs boson mass and the steadily amassing convincing statistics by the CMS, ATLAS, CDF and D\O~Collaborations of a physical 120 GeV Higgs boson, is certainly cause for reflection on whether CMS and ATLAS are indeed accumulating ${\cal F}$-$SU(5)$ supersymmetry events. We were thus motivated to ascertain the anticipated CMS and ATLAS event profile for the next significant milestone luminosity of 5 $fb^{-1}$, projected to be attainable by the conclusion of 2011. The baseline for our prediction remains our preferred gaugino mass of $M_{1/2}$=518 GeV, which fittingly generates a five standard deviation signal for the CMS experiment at 5 $fb^{-1}$ for events with greater than nine jets, the universally adopted standard signal-to-background discovery ratio, and a less substantial four standard deviation signal for the ATLAS experiment at 5 $fb^{-1}$ for $p_T >$ 80 GeV and $H_{\rm T}^{\rm miss}/\sqrt{H_{\rm T}} \ge 3.5$.

For high-energy physicists, these are certainly interesting times. The realization of a forty-year quest for supersymmetry may actually be on the proverbial doorstep of the LHC. If our bold prediction so presented here is physically observed, then the high-energy physics community in the year 2011 could finally enjoy the manifestation of the first categorical evidence of supersymmetry in our Universe.


\begin{acknowledgments}
We thank Michael Koratzinos for very helpful discussions,
which spurred certain of the new investigations in this report.

This research was supported in part 
by the DOE grant DE-FG03-95-Er-40917 (TL and DVN),
by the Natural Science Foundation of China 
under grant numbers 10821504 and 11075194 (TL),
by the Mitchell-Heep Chair in High Energy Physics (JAM),
and by the Sam Houston State University
2011 Enhancement Research Grant program (JWW).
We also thank Sam Houston State University
for providing high performance computing resources.
\end{acknowledgments}


\bibliography{bibliography}

\end{document}